\documentclass[aps,rmp,reprint,amsmath,amssymb,graphicx,longbibliography]{revtex4-1}

\usepackage{bm}
\usepackage{graphics,epsfig}
\usepackage{ulem}
\usepackage{wasysym}

\begin{document}

\begin{flushright}
{\small ~\\
TUM-HEP-1169/18
}
\end{flushright}

\title{Axion Stars}


\author{Eric Braaten}
\affiliation{Department of Physics,
         The Ohio State University, Columbus, OH\ 43210, USA}
\author{Hong Zhang}
\affiliation{Physik Department T31, Technische Universit\"at M\"unchen,
D-85748 Garching, Germany}

\date{\today{}}

\begin{abstract}
The particle that makes up the dark matter of the universe could be an {\it axion} or axion-like particle. A collection of  axions can condense into a bound Bose-Einstein condensate called an  {\it axion star}. It is possible that a significant fraction of the axion dark matter is in the form of axion stars. This would make some efforts to identify the axion as the dark matter particle more challenging, but it would also open up new possibilities. We summarize the basic properties of isolated axion stars,  which can be gravitationally bound or bound by self-interactions. Axions are naturally described by a relativistic field theory with a real scalar field, but low-energy axions can be described more simply by a classical nonrelativistic effective field theory with a complex scalar field.
\end{abstract}


\maketitle

\tableofcontents

\section{INTRODUCTION}
\label{sec:Intro}

The {\it QCD axion} is one of the best motivated candidates for the particle that makes up the dark matter of the universe, because its existence would reveal the solution to the strong $CP$ problem of QCD. (For a recent review, see Ref.~\cite{Kim:2008hd}.) The QCD axion is a spin-0 particle with very small mass and extremely weak self-interactions as well as extremely weak interactions with Standard Model particles. One might therefore expect the physics of axion dark matter to be relatively simple.

The reason axion dark matter is not so simple is that axions are identical bosons, and the axion states in dark matter may have extremely high occupation numbers. The axions can therefore form  a {\it Bose-Einstein condensate} (BEC), whose collective behavior can be quite different from an ideal gas of bosons. The axion BEC can form gravitationally bound configurations called {\it axion stars}, and it can also form self-bound configurations called {\it axitons}. If a significant fraction of the axion dark matter is in such bound configurations, it could dramatically affect experimental searches for axion dark matter.

We present a review of bound configurations of the axion BEC. Axion stars and axitons could be observed through their interactions with other astronomical objects, such as neutron stars. However we focus in this review on the simpler problem of isolated axion stars. Understanding their properties is a prerequisite for understanding some aspects of their interactions with other objects.

There are strong constraints on the most important parameters that describe the QCD axion  \cite{Kim:2008hd}. The window for the axion mass has been reduced to within one or two orders of magnitude of $10^{-4}$~eV. It should therefore be possible to understand the behavior of QCD axion dark matter in enough detail to definitively confirm its existence or rule it out. Understanding bound configurations of the axion BEC may be essential for this effort.

Axion can refer more generally to any light spin-0 particle with a periodic self-interaction potential. There are motivations from string theory for a large number of axions with masses ranging over tens of orders of magnitude, a possibility referred to as the {\it axiverse} \cite{Arvanitaki:2009fg}. There are also astrophysical motivations for a dark-matter particle that is an extremely light boson with  mass of roughly $10^{-22}$~eV \cite{Hui:2016ltb}. We focus in this review on the QCD axion, but we present our results whenever possible in a form that can be applied to other axion-like particles.

We begin the review by describing the  relativistic quantum field theory for the real scalar field of the axion. We summarize some basic features of the evolution of axion dark matter in the early universe. We then describe a nonrelativistic effective field theory with a complex scalar field that provides a simpler description of the nonrelativistic axions that make up dark matter. We discuss bound configurations of the axion BEC, first axitons and then axion stars. We conclude the review by discussing some theoretical issues involving isolated axion stars.

\section{AXION FIELD THEORY}
\label{sec:RaxionFT}

At momentum scales below the axion decay constant $f_a$, the axion can be described by a real Lorentz-scalar field in a relativistic quantum field theory.

\subsection{Fundamental Theory}
\label{sec:funFT}

The most compelling solution of the strong $CP$ problem of QCD is the {\it Peccei-Quinn mechanism}, which involves an anomalous $U(1)$ symmetry of a quantum field theory for physics beyond the Standard Model  \cite{Peccei:1977hh}. The spontaneous breaking of the Peccei-Quinn  $U(1)$ symmetry implies the existence of a spin-0 particle called the {\it axion} \cite{Weinberg:1977ma,Wilczek:1977pj}. The explicit breaking of the $U(1)$ symmetry from the chiral anomaly of QCD implies that the axion is a pseudo-Goldstone boson with nonzero mass $m_a$.

The fundamental quantum field theory for the axion is a renormalizable extension of the Standard Model in which the Peccei-Quinn symmetry is spontaneously broken by the ground state of a complex Lorentz-scalar field. The minima of its potential form a circle whose radius $f_a$ is called the {\it axion decay constant}. At momentum scales of order  $f_a$, the axion field can be identified with the Goldstone mode corresponding to excitations of the scalar field along that circle.

At momentum scales much smaller than $f_a$, the axion can be described by an elementary  real Lorentz-scalar field $\phi(x)$ with a shift symmetry: the fields $\phi(x)$ and $\phi(x)+ 2 \pi f_a$ represent the same physical configuration. The axion field has couplings to gauge fields that are determined by the anomaly, and it has derivative couplings to the other fields of the Standard Model. At momentum scales below the weak scale, which is about 100~GeV, the terms in the effective Lagrangian that couple the axion to the Standard Model fields reduce to 
\begin{equation}
\frac{\alpha_s}{8\pi f_a} \phi  \, G^a_{\mu \nu} \tilde G^{a\mu \nu} 
+ \frac{c_{\gamma 0} \alpha}{8\pi f_a}\phi \, F_{\mu \nu} \tilde F^{\mu \nu} 
+ \frac{1}{2f_a}J^\mu \partial_\mu \phi ,
\label{L-axion1}
\end{equation}
where $G^a_{\mu \nu}$ and $F_{\mu \nu}$ are the field strengths for QCD and QED, $\tilde G^a_{\mu \nu} = \frac12 \epsilon_{\mu \nu \lambda \sigma} G^{a  \lambda \sigma}$ and $\tilde F_{\mu \nu}$ are the corresponding dual field strengths, and $J^\mu$ is a linear combination of axial-vector quark currents that depends on the axion model. The coefficient $c_{\gamma 0}$ in the  $F_{\mu \nu} \tilde F^{\mu \nu}$ term in Eq.~\eqref{L-axion1} is also model dependent. The QCD field-strength term in Eq.~\eqref{L-axion1} is proportional to the topological charge density $\alpha_s G^a_{\mu \nu} \tilde G^{a\mu \nu}/8\pi$. The quantization of the QCD topological charge in the Euclidean field theory guarantees consistency with the shift symmetry of $\phi$.

In the original axion model of Peccei and Quinn, the axion decay constant $f_a$ was chosen comparable to the vacuum expectation value of the Higgs field, which is about 250 GeV. This model was quickly ruled out by the nonobservation of the axion particle in high energy physics experiments. Models with an ``invisible axion'' that are not easily ruled out by high energy physics experiments were subsequently constructed, including the KSVZ model \cite{Kim:1979if,Shifman:1979if} and the DFSZ model \cite{Dine:1981rt,Zhitnitsky:1980tq}. Astrophysical constraints from the cooling of stars by the emission of axions provide a lower bound on the axion decay constant: $f_a \gtrsim 3\times 10^9~{\rm GeV}$. The cosmological constraint that the production of axions in the early universe does not overclose the universe provides a loose upper bound: $f_a \lesssim 10^{12}~{\rm GeV}$.

\subsection{Real Scalar Field Theory}
\label{sec:realFT}

At momentum scales below the confinement scale of QCD, which is about 1~GeV, the self-interactions of axions from their couplings to the gluon field in Eq.~\eqref{L-axion1} can be described by a potential $V(\phi)$. The Lagrangian for the axion field has the form
\begin{equation}
{\cal L} = \tfrac{1}{2}\partial_\mu \phi \partial^\mu \phi - V(\phi) .
\label{L-phi}
\end{equation}
The invariance of the Lagrangian under the shift symmetry $\phi(x) \to \phi(x)+ 2 \pi f_a$ requires the axion potential $V(\phi)$ to be a periodic function of $\phi$:
\begin{equation}
V(\phi) =  V(\phi + 2 \pi f_a) .
\label{V-periodic}
\end{equation}
The Lagrangian is also  invariant under the  $Z_2$ symmetry $\phi(x) \to - \phi(x)$, which requires  $V(\phi)$ to be an even function of $\phi$. The energy density  is given by the Hamiltonian:
\begin{equation}
{\cal H} = \tfrac{1}{2} {\dot \phi}^2
 + \tfrac{1}{2} \bm{\nabla} \phi \cdot \bm{\nabla} \phi + V(\phi) ,
\label{H-phi}
\end{equation}
where $\dot \phi = \partial\phi/\partial t$ and $\nabla_i \phi = \partial\phi/\partial x^i$.

Since the axion potential  $V(\phi)$ is an even function of $\phi$, it can be expanded in powers of $\phi^2$. We choose $V(\phi)$ to have a minimum of 0 at $\phi = 0$: $V(0) =  0$. The quadratic term in the expansion determines the axion mass: $V''(\phi =0) = m_a^2$.  The expansion of  $V(\phi)$ to higher orders in $\phi$ determines the coupling constants for axion self-interactions. We define dimensionless coupling constants $\lambda_{2n}$ by using the mass $m_a$ and the decay constant $f_a$ to set the scales:
\begin{equation}
V(\phi) = \frac12 m_a^2 \phi^2
+ (m_a f_a)^2 \sum_{n=2}^{\infty} \frac{\lambda_{2n}}{(2n)!} \left( \frac{\phi}{f_a} \right)^{2n}.
\label{V-series}
\end{equation}
In reasonable axion models, the dimensionless coupling constants $\lambda_{2n}$ have natural values of order 1. The parametrization of $V(\phi)$ in Eq.~\eqref{V-series} then implies that $m_a^2/f_a^2$ is a quantum-loop factor. If this factor is small, every additional quantum loop is suppressed by an additional factor of $m_a^2/f_a^2$.

The cross section for the elastic scattering of two axions in the low-energy limit can be expressed as $\sigma = 8 \pi a^2$, where $a$ is the S-wave scattering length:
\begin{equation}
a =  (\lambda_4/32 \pi) m_a/ f_a^2.
\label{a-fa}
\end{equation}
The $Z_2$ symmetry of $V(\phi)$ implies that the number of axions in a scattering reaction is conserved modulo 2. The total number of axions is not conserved.

The equation of motion following from the Lagrangian in Eq.~\eqref{L-phi} is
\begin{equation}
\ddot \phi  = \bm{\nabla}^2 \phi - V'(\phi) .
\label{phi-EOM}
\end{equation}
The simplest periodic spherically symmetric solutions $\phi(r,t)$  can be expanded as an odd cosine series in the time $t$:
\begin{equation}
 \phi(r,t) = \sum_{n=0}^\infty  \phi_{2n+1}(r ) \, \cos\big((2n+1) \omega t\big).
\label{phi-cosine}
\end{equation}
Eq.~\eqref{phi-EOM} reduces to an infinite set of coupled equations for the harmonics $ \phi_{2n+1}(r )$. If the field $ \phi(r,t)$ always remains sufficiently small compared to $f_a$, it can be approximated by the first term in the cosine expansion:
\begin{equation}
 \phi(r,t) \approx  \phi_1(r ) \, \cos(\omega t).
\label{phi-harmonic}
\end{equation}
The {\it harmonic approximation} is obtained by inserting this expression into Eq.~\eqref{phi-EOM} and then dropping all the higher harmonics in $V'(\phi) $.

\subsection{Axion Potential} 

\begin{figure}[t]
\centerline{ \includegraphics*[width=8cm,clip=true]{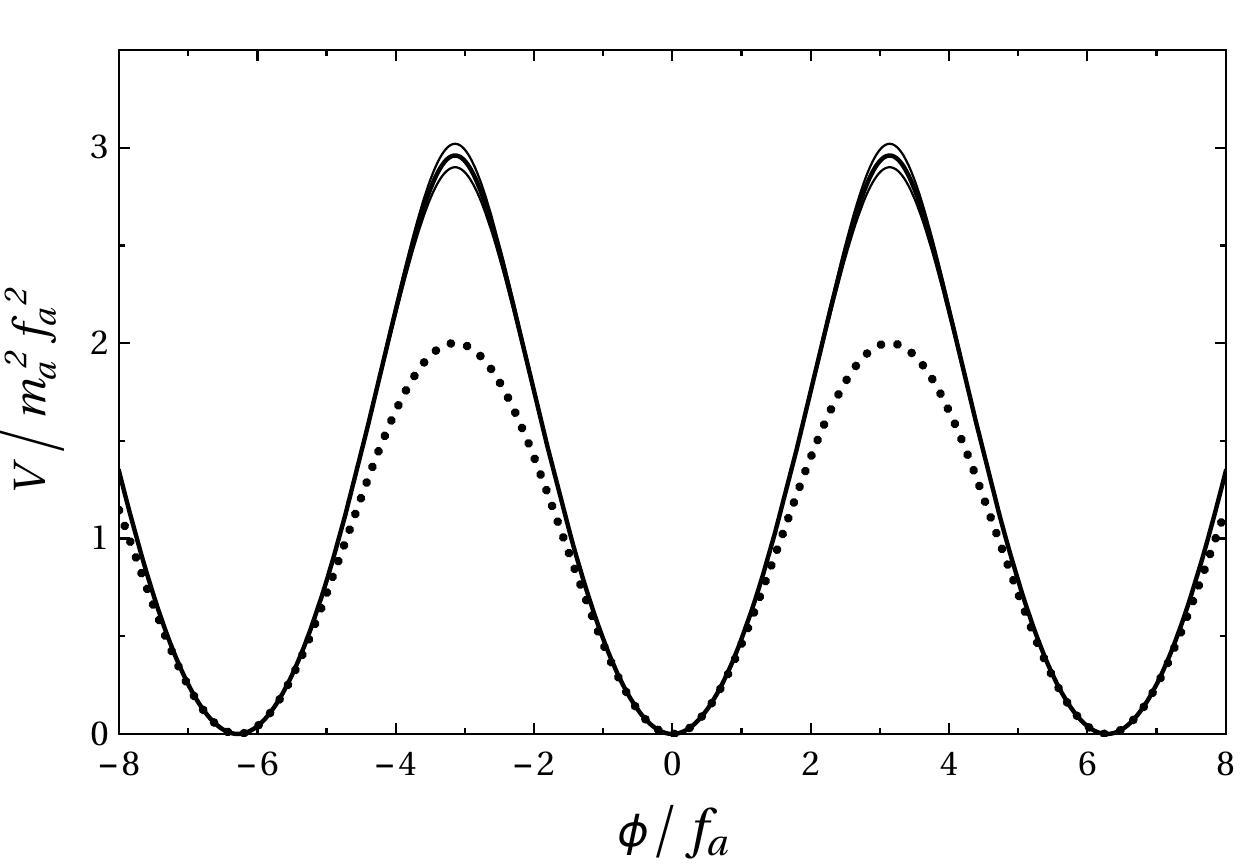} }
\vspace*{0.0cm}
\caption{Axion potential $V$ as a function of $\phi$: chiral potential for $z=0.48$ (thick solid curve) and for $z=0.45$ and 0.51 (higher and lower thin solid curves) and instanton potential (dotted curve).}
\label{fig:potential}
\end{figure}

The potential $V(\phi)$ for the axion field is determined by nonperturbative effects in QCD. A systematically improvable approximation for $V(\phi)$ can be derived from a chiral effective field theory for the light pseudoscalar mesons of QCD and the axion \cite{diCortona:2015ldu}. The leading order analysis of the chiral effective field theory for pions and the axion gives the {\it chiral  potential} \cite{DiVecchia:1980yfw}:
\begin{equation}
V(\phi) =(m_\pi f_\pi)^2
 \Bigg(\!1- \frac{\sqrt{ 1+z^2+2z  \cos (\phi/f_a)}}{1+z} \Bigg),
\label{V-chiral}
\end{equation}
where  $z = m_u/m_d$  is the ratio of the up and down quark masses. The prefactor is determined by  the mass $m_\pi = 135.0$~MeV and the decay constant  $f_\pi = 92.2$~MeV of the pion, which are related to $m_a$ and $f_a$ by
\begin{equation}
m_\pi f_\pi  = \frac{1+z}{\sqrt{z}} m_a f_a .
\label{mfa-mfpi}
\end{equation}
The numerical value of the quark mass ratio determined by a next-to-leading order analysis in the chiral effective field theory  is $z = 0.48(3)$ \cite{diCortona:2015ldu}. The chiral potential is illustrated in Fig.~\ref{fig:potential}.

The power series  in Eq.~\eqref{V-series} for the chiral potential defines dimensionless coupling constants $\lambda_{2n}$. The dimensionless coupling constant for the $4-$axion vertex is
\begin{equation}
\lambda_4 = - \frac{1-z+z^2}{(1+z)^2}.
\label{lambda4}
\end{equation}
For $z=0.48(3)$, its value is $\lambda_4 = -0.343(15)$. The resulting negative scattering length $a$ in Eq.~\eqref{a-fa} implies that axion pair interactions are attractive.

The product of $m_a$ and $f_a$ determined by a next-to-leading order analysis in the chiral effective field theory is  \cite{diCortona:2015ldu}
\begin{equation}
m_a  f_a = \big[ 75.5(5)~{\rm MeV} \big]^2.
\label{mafa}
\end{equation}
Given the upper and lower bounds on $f_a$ from cosmology and  astrophysics, the allowed mass range for the QCD axion is between $6 \times 10^{-6}$~eV and $2 \times10^{-3}$~eV. When giving the numerical value of a quantity that depends on $m_a$, we will often express the mass in the form
\begin{equation}
m_a  = 10^{-4 \pm 1} \mathrm{eV}.
\label{ma}
\end{equation}
The central value $-4$ of the exponent gives a mass near the middle of the allowed region of $m_a$ on a log scale. The $\pm1$ in the exponent should not be interpreted as an error bar, but simply as a device for specifying the dependence of a quantity on $m_a$. For example, the quantum-loop suppression factor for the QCD axion is
\begin{equation}
m_a^2/f_a^2  = 3 \times 10^{-48 \pm 4}.
\label{ma/fa}
\end{equation}
Its tiny value indicates that axions are very well described by classical field theory.

The model for the axion potential that has been used in most phenomenological studies of the axion is the {\it instanton potential}:
\begin{equation}
V(\phi)  =  (m_a f_a)^2 \big[ 1 - \cos(\phi/f_a)\big] .
\label{V-instanton}
\end{equation}
It has been derived using a dilute instanton gas approximation \cite{Peccei:1977ur}, which is not systematically improvable. The field theory defined by the Lagrangian in Eq.~\eqref{L-phi} with the instanton potential is called the {\it sine-Gordon model}. The instanton potential can be obtained from the chiral potential in Eq.~\eqref{V-chiral} by taking the limit $z\to 0$ with $m_a$ fixed. Its dimensionless coupling constants are $\lambda_{2n} = (-1)^{n+1}$. The prediction $\lambda_4 = -1$ is about a factor of 3 larger than the value  in  Eq.~\eqref{lambda4} from the chiral potential. The instanton potential is compared to the chiral potential in Fig.~\ref{fig:potential}. The potentials have the same curvature at the minima, but the amplitude of the oscillation for the instanton potential is about 2/3 that for the chiral potential with $z = 0.48$.

\subsection{Coupling to Photons} 

At momentum scales below the confinement scale of QCD, the term in the Lagrangian for the coupling of the axion to the electromagnetic field is
\begin{equation}
{\cal L}_{\rm em} = 
\frac{c_\gamma \alpha }{8\pi f_a} \phi \, F_{\mu \nu} \tilde F^{\mu\nu} .
\label{L-axionEM}
\end{equation}
The coefficient $c_\gamma$ differs from the model-dependent coefficient $c_{\gamma 0}$ in Eq.~\eqref{L-axion1} by a term that comes from a chiral transformation of the light quark fields. Its magnitude is roughly 1 in simple models. For example, $c_\gamma = -1.95$ for the simplest KSVZ model \cite{Kim:1979if,Shifman:1979if}. The decay rate of the axion into two photons is
\begin{equation}
\Gamma_a = 
\frac{c_\gamma^2 \alpha ^2 m_a^3}{256 \pi^3 f_a^2}.
\label{Gamma}
\end{equation}
In the simplest KSVZ model with $m_a =10^{-4\pm1}$~eV, the axion decay rate is $\Gamma_a = 8 \times 10^{-60 \pm 5}$~eV. The axion lifetime is $3 \!\times\! 10^{36 \mp 5}$~years. This is tens of orders of magnitude larger than the age of the universe, which is about $10^{10}$~years.

\subsection{General Relativity} 

General relativity provides a fundamental description of the gravitational interactions of axions. If the Lagrangian for $\phi$ in the absence of gravity is given in Eq.~\eqref{L-phi}, the action for $\phi$ and the space-time metric tensor $g_{\mu\nu}$ is
\begin{equation}
S = \int \!\!d^4x \sqrt{-g}
\left[ \frac{1}{2}  g^{\mu\nu}\partial_\mu\phi \partial_\nu\phi - V(\phi) - \frac{1}{16 \pi G} R \right],
\label{S-axionGR}
\end{equation}
where $g^{\mu\nu}$ is the inverse of the metric tensor, $g$ is its determinant, $R$ is the Ricci scalar, and $G$ is Newton's gravitational constant. Note that the first derivative of a scalar field is equal to its covariant derivative: $\partial_\mu \phi = D_\mu \phi$. 

The classical field equations from varying the action in Eq.~\eqref{S-axionGR} with respect to the field $\phi$ and the metric tensor $g_{\mu \nu}$ can be expressed as
\begin{subequations}
\begin{eqnarray}
&& g^{\mu\nu}D_\mu \partial_\nu\phi + V'(\phi) = 0,
\label{EOM-axion}
\\
&&T^{\mu\nu} = \frac{1}{8 \pi G} \left( R^{\mu\nu} - \tfrac12 R g^{\mu\nu} \right),
\label{EOM-Einstein}
\end{eqnarray}
\label{EOM-gravity}%
\end{subequations}
where $T^{\mu\nu}$ is the stress tensor for the scalar field and $R_{\mu\nu}$ is the Ricci tensor. If the scalar field has no self-interactions, the derivative of the potential is $V' = m_a^2 \phi$ and Eqs.~\eqref{EOM-gravity} are  the {\it Einstein-Klein-Gordon equations}.

The physics of the QCD axion involves the tiny  quantum-loop factor in Eq.~\eqref{ma/fa}. If gravity is relevant, there is another tiny number that involves Newton's  constant:
\begin{equation}
G f_a^2  =  2 \times 10^{-17\mp2}.
\label{Gf^2}
\end{equation}
The corresponding number determined by the axion mass is even tinier: $G m_a^2  =  7 \times 10^{-65\pm2}$. For gravity to be important, this tiny number must be compensated by a large number, such as the number of axions.

The tiny number $G f_a^2$ may justify the {\it linearized gravity approximation} in which Eqs.~\eqref{EOM-gravity} are linearized in fluctuations of the metric tensor around the Minkowski metric $\eta_{\mu \nu} = \mathrm{diag}(+1,-1,-1,-1)$. We set $g_{\mu \nu} = \eta_{\mu \nu}+h_{\mu \nu}$ in Eq.~\eqref{EOM-axion} and on the right side of  Eq.~\eqref{EOM-Einstein}, we set $g_{\mu \nu} = \eta_{\mu \nu}$ on the left side of Eq.~\eqref{EOM-Einstein}, and we keep only terms up to first order  in $h_{\mu \nu}$. The linearized gravity approximation should remain valid provided the system is sufficiently far from forming a black hole. The Schwarzchild radius for a black hole of mass $M$ is $2GM$. If $M(r)$ is the mass contained inside the radius $r$, the condition for the validity of the linearized approximation is $r \gg  2 \,G\,M(r)$ for all $r$. In practice, it may be sufficient for $r$ to be larger than $2 GM(r)$ by a factor of a few.

For spherically symmetric configurations, the fluctuation in the metric tensor can be expressed as
\begin{equation}
h_{\mu\nu}  =  2
\begin{pmatrix}
\Phi+3\chi & 0 \\
0 & ( \Phi - \chi) 1\!\!1 
\end{pmatrix},
\label{eq:h-phi,chi}
\end{equation}
where $\Phi(r,t)$ is the conventional gravitational potential and $\chi(r,t)$ is another gravity potential. The form of Eqs.~\eqref{EOM-gravity} with linearized gravity and spherical symmetry is such that $\phi(r,t)$ can be expanded as an odd cosine series in the time $t$, as in Eq.~\eqref{phi-cosine}, while $\Phi(r,t)$ and $\chi(r,t)$ can be expanded as even cosine series.

If the second gravity potential  $\chi$ is ignored and if time derivatives of $\Phi$ are negligible compared to its  gradients, the equations of motion in Eqs.~\eqref{EOM-gravity} reduce to
\begin{subequations}
\begin{eqnarray}
\ddot \phi - 4 \dot \Phi \dot \phi &=& 
  (1+4 \Phi)  \bm{\nabla}^2 \phi - (1+2 \Phi)V'(\phi)  ,
\label{Einstein-phi}%
\\
\bm{\nabla}^2 \Phi &=& 4 \pi G\,  \mathcal{H}.
\label{Einstein-Phi}
\end{eqnarray}
\label{Einstein-phiPhi}%
\end{subequations}
The Hamiltonian  $\mathcal{H}$ in Eq.~\eqref{H-phi} acts as the source of the gravitational potential $\Phi$.

\section{AXION DARK MATTER}
\label{sec:axionDM}

Axions can be produced in the early universe with an abundance that is compatible with the observed dark matter density. A thorough review of axion cosmology has been presented in Ref.~\cite{Marsh:2015xka}. We summarize below the aspects that are most relevant to bound configurations of an axion BEC.

\subsection{Production in the Early Universe}
\label{sec:production}

The two most important mechanisms for producing axions in the early universe are the {\it vacuum misalignment mechanism} \cite{Preskill:1982cy,Abbott:1982af,Dine:1982ah} and the {\it cosmic string mechanism} \cite{Davis:1986xc,Harari:1987ht}. The cosmic string mechanism is relevant only if the spontaneous breaking of the Peccei-Quinn symmetry occurs after inflation.

In the early universe, the quantum fields are in thermal equilibrium at a temperature $T$ that decreases as the universe expands. The metric tensor of the expanding flat  universe can be expressed as $g_{\mu \nu} = \mathrm{diag}(1,-R^2,-R^2,-R^2)$, where $R(T)$ is the  temperature-dependent scale factor that is equal to 1 at the present temperature of the cosmic microwave background. After $T$ decreases to below the scale of the axion decay constant $f_a$, the axion can be described by the real scalar field $\phi(x)$. The time evolution of the classical axion field is described by
\begin{equation}
\ddot \phi + 3 \frac{\dot R(T)}{R(T)}\, \dot \phi 
=\frac{1}{R^2(T)}\bm{\nabla}^2 \phi  - V'(\phi,T) ,
\label{phi-t:V}
\end{equation}
where $V'$ is the derivative with respect to $\phi$ of the temperature-dependent axion potential,

The axion potential $V(\phi,T)$ rises from the explicit breaking of the Peccei-Quinn symmetry by the chiral anomaly of QCD. The square of the temperature-dependent axion mass $m_a(T)$ is the QCD topological susceptibility $\chi(T)$. In the high temperature limit, $V(\phi,T)$ can be approximated by the instanton potential in Eq.~\eqref{V-instanton} with a temperature-dependent mass $m_a(T)$ that increases roughly as $T^{-4}$ as $T$ decreases  \cite{Gross:1980br}. At temperatures well below the QCD scale, $V(\phi,T)$ can be determined from a chiral effective field theory of pions and the axion \cite{diCortona:2015ldu}. For $T$ below about 100~MeV, $V(\phi,T)$ reduces to the chiral potential in Eq.~\eqref{V-chiral}. At intermediate temperatures comparable 1~GeV, $V(\phi,T)$ can be calculated nonperturbatively using lattice gauge theory. The topological susceptibility $\chi(T)$ has recently been calculated using lattice QCD with dynamical quark fields \cite{Bonati:2015vqz,Borsanyi:2016ksw,Petreczky:2016vrs,Trunin:2015yda}. The behavior of $V(\phi,T)$ near its maxima has received much less attention.

When $T$ is orders of magnitude above the QCD scale of about 1~GeV, the $V'$ term in Eq.~\eqref{phi-t:V} is negligible. The Hubble friction term proportional to $\dot \phi$ causes $\phi(x)$ to relax to a time-independent value $\bar \phi$. There is no energetically preferred value of $\phi$, so $\bar \phi$ varies slowly across the universe almost everywhere. Inside any circle around which  $\bar \phi$ changes continuously from 0 to  $2 \pi f_a$, there must be a topological defect called a {\it cosmic string}. The cosmic string is a narrow tube with width of order $1/f_a$ inside which the Peccei-Quinn symmetry remains unbroken. As the universe continues to expand and cool, the network of cosmic strings evolves, with small closed loops shrinking and disappearing, long cosmic strings becoming straighter, and cosmic strings crossing and reconnecting. In all these processes, axions are radiated. When $T$ decreases to below the QCD scale, the remaining cosmic strings decay into axions. Most of the axions from this {\it cosmic string mechanism} are relativistic and  incoherent when they are produced. The subsequent Hubble expansion makes them highly nonrelativistic and gives them huge occupation numbers.

As the temperature $T$ decreases towards the  QCD scale, the $V'$ term in Eq.~\eqref{phi-t:V}  becomes increasingly important. The energetically preferred values of the axion field $\phi$ are the minima of $V(\phi,T)$ at $n(2 \pi f_a)$, where $n$ is an integer. In a region where the axion field has relaxed to a value $\bar \phi$ between $-\pi f_a$ and $+\pi f_a$, the field begins to oscillate around 0 with amplitude $|\bar \phi|$. As the temperature decreases, the Hubble expansion rate $\dot R/R$ decreases roughly as $T^2$. The oscillations are at first damped by the Hubble friction term in Eq.~\eqref{phi-t:V}, but when its effects become negligible, the field continues to oscillate with a smaller amplitude $\bar \phi$ that varies slowly in space. Such an oscillation can be interpreted as a BEC of axions with number density $m_a \bar \phi^2$. The {\it vacuum misalignment mechanism} is the production of axions in the form of these oscillations of the axion field. These axions are coherent, highly nonrelativistic, and have huge occupation numbers.

The vacuum misalignment mechanism and the cosmic string mechanism have traditionally been considered as two independent mechanisms whose contributions to axion production must be added. An updated calculation of the vacuum misalignment mechanism is given in Ref.~\cite{Bae:2008ue}. A recent calculation of the cosmic string mechanism is presented in Ref.~\cite{Hiramatsu:2010yu}. The string tension of the cosmic strings depends logarithmically on the large ratio of $f_a$ to the inverse of the Hubble length $1/R(t)$. Most previous numerical simulations have required extrapolations in the string tension by about an order of magnitude. Klaer and Moore have recently pointed out that in numerical simulations that take into account the large string tension, the vacuum misalignment mechanism must be considered simultaneously \cite{Klaer:2017ond}. Further progress on numerical simulations of axion production should allow definitive predictions for the production of the QCD axion in the early universe \cite{Gorghetto:2018myk,Kawasaki:2018bzv}.

\subsection{Axion Miniclusters}
\label{sec:axionMC}

When the temperature of the early universe is comparable to the QCD scale of about 1~GeV, regions in which $\bar \phi$ is farther from the minima of $V(\phi,T)$ have larger axion energy density. Hogan and Rees pointed out that the overdense regions can become gravitationally bound, and they can decouple from the Hubble expansion of the universe \cite{Hogan:1988mp}. They referred to these gravitationally bound systems of axions as {\it axion miniclusters}. When the axion dark matter evolves into the halos of galaxies, the axion miniclusters may become localized regions in which the mass density is many orders of magnitude larger than the local dark matter mass density.
 
The distribution of the mass $M$  for axion miniclusters has been calculated using various simplifying assumptions \cite{Fairbairn:2017sil,Enander:2017ogx}. The temperature-dependent mass of the axion was taken into account, but the periodicity of $V(\phi,T)$ as a function of $\phi$ was ignored. For the QCD axion with mass $10^{-4}$~eV, the peak in the distribution at the time of matter-radiation equality is near $10^{-14} M_\odot$, where $M_\odot$ is the mass of the sun \cite{Enander:2017ogx}. As the universe evolves further, the distribution for $M$ broadens. The value of $M$ at the peak decreases, while the distribution expands to higher $M$ from the merger of miniclusters into more massive miniclusters \cite{Fairbairn:2017sil}.

There do not seem to be any quantitative theoretical predictions of the fraction $f_\mathrm{mc}$ of the axion dark matter in the form of axion miniclusters. Gravitational microlensing has been used to place an upper bound on $f_\mathrm{mc}$ \cite{Fairbairn:2017dmf}. For the QCD axion with mass $m_a = 10^{-4\pm1}$~eV, the bound is $f_\mathrm{mc}< 0.08 \times 10^{\pm 0.12}$. 

Kolb and Tkachev studied the time evolution of axion miniclusters as the temperature decreases to below the QCD scale  by solving Eq.~\eqref{phi-t:V} for the axion field in the expanding flat universe \cite{Kolb:1993zz,Kolb:1993hw}. For the potential $V(\phi,T)$, they used the instanton potential in Eq.~\eqref{V-instanton} with a temperature-dependent axion mass $m_a(T)$. Their initial configurations had many local peaks in the axion energy density. As time evolves, the lower peaks remain almost unchanged, consistent with being frozen by  Hubble friction. However the higher peaks, which  can be identified as axion miniclusters, become smaller in size and roughly spherically symmetric. The axion field $\phi(r,t)$ oscillates rapidly with angular frequency near $m_a$. The amplitude $\phi_0$ at the center is a  substantial fraction of $\pi f_a$, which is the value of $\phi$ where the axion potential is maximum. Kolb and Tkachev studied the subsequent time evolution of individual peaks by solving Eq.~\eqref{phi-t:V} with spherical symmetry. The time evolution has three stages in which the amplitude $\phi_0$ at the center has different behavior:
(1) $\phi_0$ is a substantial fraction of $\pi f_a$,
(2) there are multiple cycles in which  $\phi_0$ grows to near $\pi f_a$ for a while and then  suddenly collapses, 
(3)  $\phi_0$ is much smaller than $\pi f_a$.
Kolb and Tkachev referred to these 3-stage localized axion field configurations as  {\it axitons}. The possibility of producing axitons has not been taken into account in most subsequent studies of axion miniclusters.

\subsection{Thermalization}
\label{sec:axBEC}

The vacuum misalignment mechanism produces axions that are already in a BEC. The cosmic string  mechanism produces incoherent axions with very large occupation numbers. If there is a thermalization mechanism that can bring these axions into coherence, they can also become a BEC. When the temperature is comparable to the QCD scale, the thermalization rate due to $2 \to 2$ axion scattering is comparable to the Hubble expansion rate \cite{Sikivie:2010bq}. The time scale required for the formation of a condensate is much shorter than that for reaching thermodynamic equilibrium \cite{Berges:2014xea}. Thus, as the temperature of the expanding universe decreases below the QCD scale, the nonrelativistic axions from both mechanisms should form a locally homogeneous BEC.

As the temperature $T$ of the universe continues to decrease, the axion BEC will evolve in accord with classical field equations. A sufficiently effective thermalization mechanism would drive the BEC toward the lowest-energy states that are accessible. When $T$ is well below the QCD scale,  $2 \to 2$ axion scattering is no longer effective. However Sikivie and Yang pointed out that the gravitational scattering of axions provides a thermalization mechanism for the axion BEC at temperatures below about $100~\mathrm{eV} \times (f_a/10^{12}~\mathrm{GeV})^{1/2}$ \cite{Sikivie:2009qn}. Other investigators have obtained similar results \cite{Saikawa:2012uk,Noumi:2013zga}. Those results have also been questioned \cite{Davidson:2013aba,Davidson:2014hfa}. Sikivie and collaborators have shown that rethermalization of the axion BEC can have observable effects on the dark matter halos of galaxies \cite{ Sikivie:2010bq, Erken:2011vv, Erken:2011dz}.

Guth, Hertzberg, and Prescod-Weinstein have argued that the attractive interactions from gravity and from axion self-interactions will prevent the formation of an axion BEC with coherence length the size of a galaxy  \cite{Guth:2014hsa}. They pointed out that a locally homogeneous BEC of axions is unstable to formation of localized denser clumps of axions. The clumps could be {\it axitons}  bound by axion self-interactions (see Section~\ref{sec:BoundAxions}) or {\it axion stars}  bound by gravity (see Section~\ref{sec:GravityBound}). In the case of axion stars,  gravitational cooling provides an efficient mechanism for relaxation to a stable configuration \cite{Seidel:1993zk,Guzman:2006yc}.

\section{NONRELATIVISTIC EFFECTIVE FIELD THEORY}
\label{sec:NRaxionEFT}

At momentum scales below the axion mass $m_a$, the axion can be described most simply by a complex scalar field in a nonrelativistic effective field theory.

\subsection{Complex Field} 

Given the relativistic quantum field theory for the real Lorentz-scalar field $\phi(x)$ with the Lagrangian  in Eq.~\eqref{L-phi}, a {\it nonrelativistic effective field theory} (NREFT)  can be obtained by integrating out the scale of the mass $m_a$. The effective field theory describes particles with momenta small compared to $m_a$. It should also describe field configurations with gradients small compared to $m_a$ and with angular frequencies sufficiently close to $m_a$.

Perhaps somewhat surprisingly, the most convenient field for NREFT is a complex scalar field $\psi(\bm{r},t)$. The complex field can be identified naively with the positive-frequency component of the real scalar field with a phase factor $\exp(-i m_a t)$ removed:
\begin{equation}
\phi(\bm{r},t) \approx
\frac{1}{\sqrt{2m_a}}  \Big( \psi(\bm{r},t) \, e^{-i m_a t} +  \psi^*(\bm{r},t) \, e^{+i m_a t} \Big).
\label{phi-psi}
\end{equation}
A naive  effective Lagrangian for $\psi$ can be obtained by inserting this expression for $\phi$ into the Lagrangian  in Eq.~\eqref{L-phi} and then dropping terms with rapidly changing phase factors of the form $\exp(inm_at)$ with nonzero integer $n$. The resulting Lagrangian is
\begin{eqnarray}
\mathcal{L}_\mathrm{naive} &=&
 \frac{i}{2} \left( \psi^* \dot \psi - \dot \psi^*\psi \right)  
-\frac{1}{2m_a} \nabla \psi^* \cdot \nabla \psi
\nonumber\\
&& 
- V_\mathrm{eff}^{(0)}(\psi^*\psi) + \frac{1}{2m_a} \dot \psi^* \dot \psi,
\label{L-naive}
\end{eqnarray}
where the naive effective potential $V_\mathrm{eff}^{(0)}$ is a function of $\psi^*\psi$. The terms $\frac12\dot \phi^2$ and $-V$ in Eq.~\eqref{L-phi} give canceling mass terms $\pm \frac12 m_a \psi^*\psi$. If the axion potential $V(\phi)$ has the power series in Eq.~\eqref{V-series}, the power series for the naive effective potential is \cite{Eby:2014fya}
\begin{equation}
V_\mathrm{eff}^{(0)}(\psi^* \psi) = 
m_a^2 f_a^2 \sum_{n=2}^{\infty} \frac{\lambda_{2n}}{(n!)^2} \left(\frac{\psi^* \psi}{2m_af_a^2} \right)^n.
\label{V-naive}
\end{equation}

NREFT can be derived rigorously by starting from an exact connection between the real field $\phi(\bm{r},t)$ and a complex field $\psi(\bm{r},t)$ \cite{Namjoo:2017nia}. The real field and its time derivative $\dot \phi$, which is the momentum conjugate to $\phi$, can be expressed as
\begin{eqnarray}
\phi(\bm{r},t)  &=&
\frac{1}{\sqrt{2m_a}} \mathcal{D}
 \Big( \psi(\bm{r},t) \, e^{-i m_a t} +  \psi^*(\bm{r},t) \, e^{+i m_a t} \Big),
 \nonumber\\
\dot \phi(\bm{r},t)  &=& \frac{-im_a}{\sqrt{2m_a}} \mathcal{D}^{-1}
 \Big( \psi(\bm{r},t) \, e^{-i m_a t} -  \psi^*(\bm{r},t) \, e^{+i m_a t} \Big),
 \nonumber\\
\label{phi-psi:CT}%
\end{eqnarray}
where $\mathcal{D}= (1 - \bm{\nabla}^2/m_a^2)^{-1/4}$. These expressions are local in time, but the operators $\mathcal{D}$ and $\mathcal{D}^{-1}$ make them nonlocal in space. This transformation between the complex fields $\psi$ and $\psi^*$ and the real fields $\phi$ and $\dot\phi$ is a canonical transformation \cite{Namjoo:2017nia}. The quantum field $\phi$ satisfies canonical local equal-time commutation relations, including $[\phi(\bm{r},t), \dot \phi(\bm{r'},t)] = i\, \delta^3(\bm{r} - \bm{r'})$. Remarkably, despite the nonlocality of Eqs.~\eqref{phi-psi:CT}, they imply that $\psi$ satisfies the canonical local equal-time commutations  relations for a nonrelativistic field, including $[\psi(\bm{r},t), \psi^*(\bm{r'},t)] = \delta^3(\bm{r} - \bm{r'})$. Once the transformations in Eqs.~\eqref{phi-psi:CT} have been used to eliminate the real field $\phi(\bm{r},t)$ in favor of the complex field $\psi(\bm{r},t)$,  NREFT can in principle be derived simply by integrating out the momentum scale $m_a$ from the complex field.

\subsection{Effective Lagrangian} 

Having identified the field that describes the low-energy degrees of freedom, the Lagrangian for NREFT can be constructed more efficiently by using the matching methods of effective field theory.  The effective Lagrangian is assumed to include all possible local terms consistent with the symmetries of the original theory at low energy. The coefficients of these terms are determined by matching quantities calculated in both the original theory and the effective theory. The principles of effective field theory guarantee that low-energy observables can be reproduced in the effective theory with a systematically improvable accuracy. More specifically, if an observable involves only low energies $E$ satisfying $|E| \ll m_a$, its expansion in powers of $E/m_a$ can be reproduced. An important caveat is that NREFT may not reproduce terms with an essential singularity in $E/m_a$.

The NREFT for a real Lorentz-scalar field was first constructed in Ref.~\cite{Braaten:2015eeu}. It was called {\it axion EFT}, although it can be applied to any field theory with a real Lorentz-scalar field. The effective Lagrangian $\mathcal{L}_\mathrm{eff}$ for NREFT can be chosen to have simple linear dependence on the time derivative $\dot \psi$ \cite{Braaten:2015eeu}:
\begin{equation}
\mathcal{L}_\mathrm{eff} = 
\tfrac12 i \left( \psi^* \dot \psi - \dot \psi^*\psi \right) - \mathcal{H}_\mathrm{eff}.
\label{Leff-psi}
\end{equation}
The effective Hamiltonian density depends on the field $\psi$ and its gradients, but not on its time derivatives. It can be expressed as
\begin{equation}
\mathcal{H}_\mathrm{eff}  = 
\mathcal{T}_\mathrm{eff} + V_\mathrm{eff} + W_\mathrm{eff},
\label{Heff-psi}
\end{equation}
where $\mathcal{T}_\mathrm{eff}$ is the kinetic energy density, $V_\mathrm{eff} $ is the effective potential, which  is a function of  $\psi^* \psi$ only, and $W_\mathrm{eff}$ consists of all interaction terms that depend also on gradients of $\psi$. An $n$-body term in $\mathcal{H}_\mathrm{eff} $ has $n$ factors of $\psi$ and  $n$ factors of $\psi^*$. The kinetic energy density includes all the one-body terms:
\begin{equation}
\mathcal{T}_\mathrm{eff} = 
\frac{1}{2m_a} \nabla \psi^* \cdot \nabla \psi
- \frac{1}{8m_a^3} \nabla^2 \psi^* \,  \nabla^2 \psi
+ \ldots.
\label{Teff-psi}
\end{equation}
These terms reproduce the energy-momentum relation $E = \sqrt{m^2+p^2} -m$. The effective potential $V_\mathrm{eff}$ can be expanded in powers of $\psi^*\psi$ beginning at order $(\psi^*\psi)^2$:
\begin{equation}
V_\mathrm{eff}(\psi^* \psi) = 
 m_a^2 f_a^2 \sum_{n=2}^{\infty} \frac{v_n}{(n!)^2} \left(\frac{\psi^* \psi}{2m_af_a^2} \right)^n.
\label{Veff-series}
\end{equation}
In Ref.~\cite{Braaten:2016kzc}, $V_\mathrm{eff}$ included the 1-body term $m_a \psi^* \psi$. It is more natural to omit this term, since the effective field theory is obtained by integrating out  the momentum scale $m_a$. The terms in $W_\mathrm{eff}$ in Eq.~\eqref{Heff-psi} are $n$-body interaction terms with $n \ge 2$ that depend on gradients of $\psi$.

An alternative formulation of NREFT with a complex scalar field $ \psi$ was proposed in Ref.~\cite{Mukaida:2016hwd}. Their effective Lagrangian  includes the $\dot \psi^* \dot \psi$ term in the naive effective Lagrangian in Eq.~\eqref{L-naive}, which is quadratic in the time derivative. The effective potential also differs from $V_\mathrm{eff}$ in Eq.~\eqref{Veff-series} beginning with the $(\psi^* \psi)^3$ term. The equivalence of the two effective field theories was demonstrated in Refs.~\cite{Braaten:2016kzc,Namjoo:2017nia} by showing that they give the same T-matrix elements for low-energy scattering and by showing that the two effective Lagrangians differ by a redefinition of the complex field $ \psi(\bm{r},t)$. The equivalence requires that the effective Lagrangian for the NREFT in Ref.~\cite{Mukaida:2016hwd} include interaction terms that depend on $\dot\psi$, such as $(\psi^* \psi)^2(\psi^* \dot \psi - \dot \psi^*\psi)$ \cite{Braaten:2016kzc}.

The equation of motion for $\psi(\bm{r},t)$ in NREFT with the Lagrangian in Eq.~\eqref{Leff-psi} has infinitely many terms. A simpler equation can be obtained by omitting terms in the effective Hamiltonian that are suppressed. If we consider a configuration  in which all gradients are small compared to $m_a$, we can omit all but the first term in the kinetic energy density  in Eq.~\eqref{Teff-psi} and we can omit the term  $W_\mathrm{eff}$, which consists of interaction terms with gradients. The equation of motion reduces to
\begin{equation}
i \dot \psi=  - \frac{1}{2 m_a}\bm{\nabla}^2 \psi + V_\mathrm{eff}'(\psi^* \psi)\,  \psi,
\label{psi-EOM}
\end{equation}
where $V_\mathrm{eff}'$ is the derivative with respect to the argument $\psi^*\psi$. If the effective potential is $V_\mathrm{eff} = (2 \pi a/m_a) (\psi^*\psi)^2$, where $a$ is the S-wave scattering length, this is the {\it Gross-Pitaevskii equation}. The simplest spherically symmetric solutions of Eq.~\eqref{psi-EOM} have harmonic time dependence: 
\begin{equation}
\psi(\bm{r},t) = \psi_0(r)\, e^{+i \varepsilon_bt}.
\label{psi-periodic}
\end{equation}
If $\varepsilon_b>0$, it can be interpreted as the binding energy of a boson. Eq.~\eqref{psi-EOM} with the naive effective potential $V_\mathrm{eff}^{(0)}$ in Eq.~\eqref{V-naive} is equivalent to the harmonic approximation for the real scalar field obtained by inserting Eq.~\eqref{phi-harmonic} into Eq.~\eqref{phi-EOM} \cite{Visinelli:2017ooc}. The field $\psi_0(r)$ in Eq.~\eqref{psi-periodic} can be identified with the first harmonic $\phi_1(r)$ in Eq.~\eqref{phi-harmonic} multiplied by $\sqrt{m_a/2}$.

NREFT is especially useful when quantum loops are strongly suppressed by a tiny quantum-loop factor, such as $m_a^2/f_a^2$ in Eq.~\eqref{ma/fa}. In Ref.~\cite{Braaten:2018lmj}, the approximation to NREFT in which all quantum loops are omitted was called  {\it classical nonrelativistic effective field theory} (CNREFT). However there are also purely classical effects that are suppressed by $m_a^2/f_a^2$. It may be more convenient to define CNREFT to be the approximation to NREFT in which all effects suppressed by $m_a^2/f_a^2$ are omitted.

\subsection{Effective Potential} 

\begin{figure}[t]
\centerline{ \includegraphics*[width=8cm,clip=true]{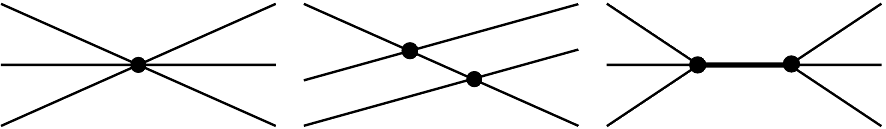} }
\vspace*{0.0cm}
\caption{The tree-level diagrams for low-energy $3 \to 3$ scattering in the relativistic axion theory. The first two diagrams are also diagrams in NREFT. In the third diagram, the thicker line indicates a virtual axion whose invariant mass is approximately $3m$.}
\label{fig:3to3tree}
\end{figure}

The dimensionless coupling constants $v_{n}$ for the $n \to n$ axion vertices are coefficients in the power series for the effective potential $V_\mathrm{eff}$ in Eq.~\eqref{Veff-series}. They can be determined to leading order in $m_a^2/f_a^2$ by matching tree-level scattering amplitudes in the relativistic theory and in NREFT in the limit where all the external 3-momenta go to 0 \cite{Braaten:2016kzc}. The coupling constant $v_2 = \lambda_4$  is obtained simply by matching the $2 \to 2$ scattering amplitude, which  is given  by the $2 \to 2$ vertex in both the relativistic theory and NREFT. The coupling constant $v_3$  can then be obtained by matching the $3 \to 3$ scattering amplitude, which is given in the relativistic theory by the sum of the 3 diagrams in Fig.~\ref{fig:3to3tree} and in NREFT   by the sum of the first 2 diagrams in Fig.~\ref{fig:3to3tree}. The coupling constant $v_4$  can be obtained by matching the $4 \to 4$ scattering amplitude. The results for these first three coefficients  are
\begin{subequations}
\begin{eqnarray}
v_2 &=& \lambda_4,  \qquad
v_3 =  \lambda_6  - (17/8) \lambda_4^2,
\label{v3}
\\
v_4 &=&  \lambda_8 - 11 \lambda_4 \lambda_6 + (125/8) \lambda_4^3.
\label{v4}
\end{eqnarray}
\label{v2345}%
\end{subequations}
The result for $v_3$ was first obtained in Ref.~\cite{Braaten:2016kzc}. The result for $v_4$ was first calculated correctly in Ref.~\cite{Braaten:2018lmj}.

\begin{figure}[t]
\centerline{ \includegraphics*[width=8cm,clip=true]{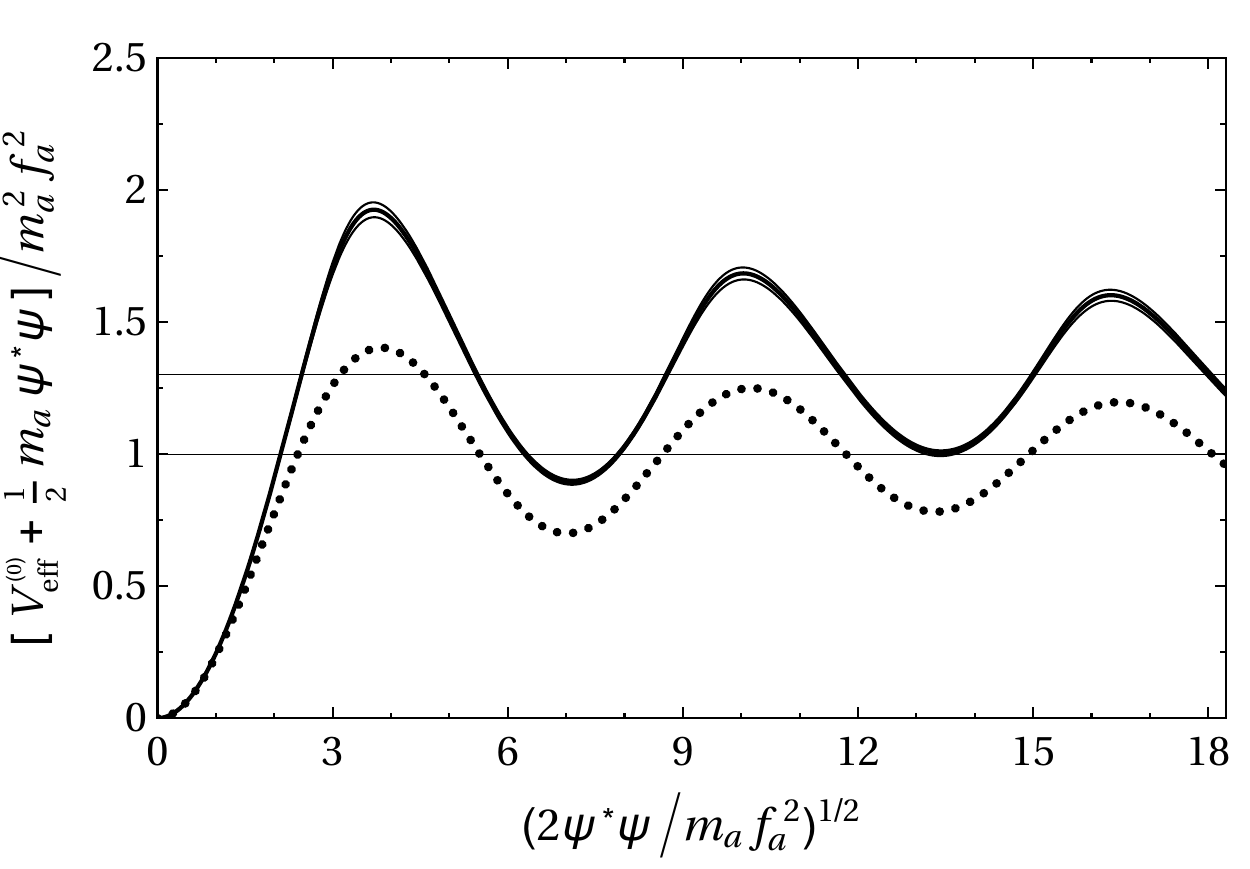} }
\vspace*{0.0cm}
\caption{Naive effective potential $V_\mathrm{eff}^{(0)}$ with $\tfrac12  m_a \psi^* \psi$ added as a function of $|\psi|$: chiral potentials for $z=0.48$ (thicker solid curve) and for $z=0.45$ and 0.51 (thinner solid curves) and instanton potential (dotted curve). The thin horizontal lines are the asymptotic values at large $|\psi|$.}
\label{fig:Vclass}
\end{figure}

Successive truncations of  the power series for the effective potential $V_\mathrm{eff}$ in Eq.~\eqref{Veff-series} define increasingly accurate approximations only if $\psi^*\psi$ is very small compared to $m_a f_a^2$. For $\psi^*\psi$ of order $m_a f_a^2$, an approximation to $V_\mathrm{eff}$  must include terms with arbitrarily high powers of $\psi^*\psi$. An example is the naive effective potential $V_\mathrm{eff}^{(0)}$ in Eq.~\eqref{V-naive}, which was first introduced in Ref.~\cite{Eby:2014fya}. The naive effective potential for the chiral potential in Eq.~\eqref{V-chiral} has a convenient integral representation  \cite{Eby:2017xrr}:
\begin{eqnarray}
V_\mathrm{eff}^{(0)}(\psi^* \psi) &=&
(m_\pi f_\pi)^2 \bigg( 1 - \frac{z}{4(1+z)^2}\hat n
\nonumber\\
&& \hspace{-2.0cm} 
-\frac{1}{1+z} \int_0^1 dt  \sqrt{ 1+z^2+2z  \cos ( \hat n^{1/2} \sin(\pi t))} \, \bigg),
\label{Veff-chiral}
\end{eqnarray}
where the dimensionless number density $\hat n$ is
\begin{equation}
\hat n = 2\psi^* \psi/(m_af_a^2).
\label{psi-hat}
\end{equation}
The naive effective potential for the instanton potential in Eq.~\eqref{V-instanton} can be expressed  analytically in terms of a Bessel function \cite{Eby:2014fya}:
\begin{equation}
V_\mathrm{eff}^{(0)}(\psi^* \psi) =
 (m_a f_a)^2 \left[ 1- \tfrac14 \hat n - J_0( \hat n^{1/2}) \right] .
\label{Veff-instanton}
\end{equation}
At large $\psi^*\psi$, the naive effective potentials  in Eqs.~\eqref{Veff-chiral} and \eqref{Veff-instanton} approach $-\tfrac12  m_a \psi^* \psi$ plus a constant that is different for the chiral potential and the instanton potential. As shown in Fig.~\ref{fig:Vclass}, the approach to the constant is through oscillations in $\hat n^{1/2}$ with decreasing amplitude.

A sequence of improved effective potentials $V_\mathrm{eff}^{(k)}$, each of which includes additional terms with all powers of $\psi^*\psi$, was proposed in Ref.~\cite{Braaten:2016kzc}. The effective potential $V_\mathrm{eff}^{(k)}$ is defined by matching tree-level scattering amplitudes in the low-momentum limit in the relativistic theory and in NREFT from diagrams with at most $k$ virtual particles. For $k=0$, this gives the naive effective potential $V_\mathrm{eff}^{(0)}$ defined by the power series in Eq.~\eqref{V-naive}. The coefficients in the power series for the first improved effective potential  $V_\mathrm{eff}^{(1)}$ were determined in Ref.~\cite{Braaten:2016kzc}. 

\subsection{Axion Loss Terms} 

The decay of an axion into two photons produces photons with energies $\frac12 m_a$ in the axion rest frame. States that include these high-energy photons cannot be described explicitly in  NREFT, but the effects of  axion decay on nonrelativistic axions  can be described by adding the anti-Hermitian term $-i (\Gamma_a/2) \psi^*\psi$  to the effective Hamiltonian density, where $\Gamma_a$ is the axion decay rate in Eq.~\eqref{Gamma}.

There are reactions that  produce relativistic axions from initial states that include only nonrelativistic axions. The simplest such reaction is $4  \to 2$ scattering, which produces two relativistic axions each with energy close to $2m_a$. States that include relativistic axions cannot be described  explicitly in NREFT. However the effects on nonrelativistic axions from the reactions that produce relativistic axions can be taken into account systematically through anti-Hermitian terms in the  effective Hamiltonian density $\mathcal{H}_\mathrm{eff}$. The highly inelastic nature of the reactions that produce relativistic axions ensures that their effects can be described by local terms in $\mathcal{H}_\mathrm{eff}$ \cite{Braaten:2016sja}.

The  imaginary part of the effective potential for NREFT can be expanded in powers of  $(\psi^*\psi)^2$ beginning at order $(\psi^*\psi)^4$:
\begin{equation}
\mathrm{Im} \,V_\mathrm{eff}(\psi^* \psi) = 
- m_a^4 \sum_{n=2}^{\infty} \frac{x_{2n}}{\big( (2n)! \big)^2} \left(\frac{\psi^* \psi}{2m_af_a^2} \right)^{2n}.
\label{ImVeff-series}
\end{equation}
The dimensionless coupling constants $x_4$ and $x_6$  were calculated to leading order in $m_a^2/f_a^2$ in Ref.~\cite{Braaten:2016dlp} by matching the square of the tree-level $2n \to2$ amplitude with vanishing incoming 3-momenta in the relativistic theory with the imaginary part of the tree-level  amplitude in NREFT. The results are
\begin{equation}
x_4 = \frac{\sqrt{3}}{64 \pi}\,\big[\lambda_6 - \lambda_4^2\big]^2,
~~~
x_6 = \frac{\sqrt{2}}{48 \pi} \,\big[\lambda_8  - \lambda_4 \lambda_6\big]^2.
\label{eq:w4,w6}%
\end{equation}
(These coefficients are  both zero for the instanton potential whose dimensionless coupling constants are $\lambda_{2n} = (-1)^{n+1}$.) The coefficient of $(\psi^* \psi)^{2n}$ in $\mathrm{Im} \,V_\mathrm{eff}$ is suppressed by a factor of $m_a^2/f_a^2$ compared to the coefficient of $(\psi^* \psi)^{2n}$ in $\mathrm{Re} \,V_\mathrm{eff}$ in Eq.~\eqref{Veff-series}. This suppression follows from the optical theorem, which implies that the square of the tree-level $2n \to2$ amplitude is proportional to the imaginary part of a sum of one-loop diagrams and must therefore be suppressed by the same factor as a quantum loop.

The Lagrangian for NREFT in Eq.~\eqref{Leff-psi} has a $U(1)$ symmetry in which the field $\psi(\bm{r},t)$ is multiplied by a phase $e^{i \alpha}$. If the Hamiltonian density  was Hermitian, the $U(1)$ symmetry  would imply conservation of a particle number: 
\begin{equation}
N = \int\!\! d^3r\,  \psi^* \psi .
\label{Naxion}
\end{equation}
This number $N$ should be interpreted as the total number of nonrelativistic axions. However the Hamiltonian density for NREFT includes anti-Hermitian terms. Its equations of motion therefore predict that $N$ decreases continually with time. The rate of decrease in $N$ is predicted to be
\begin{equation}
- \frac{d\ }{dt}N =  \Gamma_a N  
+m_a^4 \int\! d^3r\,\bigg[ \frac{x_4}{72} \left(\frac{\psi^*\psi}{2 m_a f_a^2}\right)^4  + \ldots \bigg],
\label{dN/dt-X}
\end{equation}
where $\Gamma_a$ is the decay rate of the axion into two photons  in Eq.~\eqref{Gamma}. In the second term, we have shown explicitly only the  $(\psi^*\psi)^4$ term from $4 \to 2$ reactions. The corresponding term in the energy loss rate $dE/dt$ was derived by Hertzberg using the real scalar field theory \cite{Hertzberg:2010yz}. He referred to the loss process as  ``quantum radiation''. This name is  misleading, because this is a classical loss process whose rate can be derived from tree diagrams. We will see in Section~\ref{sec:Emission} that the $4 \to 2$ reaction does not give the leading contribution to the loss rate of nonrelativistic axions. There are additional mechanisms for the emission of relativistic axions whose effects on nonrelativistic axions do not seem to be reproduced by the low-energy effective field theory NREFT. 

\subsection{Newtonian Gravity} 

A fundamental description of axions with gravity is provided by the action from general relativity in Eq.~\eqref{S-axionGR} for the real scalar field $\phi(x)$ and the metric tensor $g_{\mu\nu}(x)$. A low-energy effective field theory for  axions with gravity could be obtained by integrating out fluctuations with momenta of order $m_a$ and larger for both $\phi(x)$ and the metric tensor. The fields that describe the low-energy degrees of freedom are a complex scalar field $\psi(\bm{r},t)$ and additional fields that arise from the metric tensor. If the additional fields are identified, the effective Lagrangian could be constructed by using the matching methods of effective field theory. It would be necessary to match scattering amplitudes not only for scalars but also for gravitons.

An alternative approach would be to start from the low-energy effective field theory NREFT in the absence of gravity, whose Lagrangian is given in  Eq.~\eqref{Leff-psi}, and then introduce gravity by requiring consistency with general coordinate invariance. There is a truncation of the effective Lagrangian for NREFT that retains the most important terms in the nonrelativistic limit and has Galilean symmetry. For example, the kinetic energy density in Eq.~\eqref{Teff-psi} must be truncated after the $\nabla \psi^* \cdot \nabla \psi$ term. A general coordinate invariance  for nonrelativistic field theories with Galilean symmetry has been introduced by Son and Wingate \cite{Son:2005rv}. It might be possible to use this nonrelativistic general coordinate invariance to deduce terms in the Lagrangian for NREFT with gravity.

General relativity applied to matter systems reduces in the nonrelativistic limit to Newtonian gravity plus post-Newtonian corrections. A simple guess for the low-energy effective field theory for axions with gravity is that it reduces to NREFT for the complex field $\psi$ with Newtonian gravity. In Newtonian gravity, the space-time-dependent metric tensor reduces to a single real function: the gravitational potential $\Phi(\bm{r},t)$. If we keep only the leading term\ in the kinetic energy density in Eq.~\eqref{Teff-psi} and omit the gradient interaction term $W_\mathrm{eff}$, the effective action reduces to
\begin{eqnarray}
S  &= & \int \!\!dt \int \!\!d^3r\bigg( \frac{i}{2} ( \psi^* \dot \psi -  \dot \psi^* \psi )
 -\frac{1}{2m_a} \bm{\nabla} \psi^* \cdot \bm{\nabla} \psi
\nonumber\\
& &
 - \frac{1}{8 \pi G} \bm{\nabla} \Phi \cdot \bm{\nabla} \Phi - V_\mathrm{eff}(\psi^* \psi) - m_a \psi^* \psi\,  \Phi \bigg).
\label{S-psiPhi}
\end{eqnarray}
Since the Lagrangian does not depend on time derivatives of $\Phi$, the gravitational potential is not dynamical. Its  source  is the mass density $m_a \psi^* \psi$. The variational equations are
\begin{subequations}
\begin{eqnarray}
i \dot \psi &=&  - \frac{1}{2m_a} \bm{\nabla}^2 \psi 
+\left[ V_\mathrm{eff}'(\psi^* \psi)  +m_a \Phi \right] \psi,
\label{Newton-psi}
\\
\bm{\nabla}^2 \Phi &=& 4 \pi Gm_a \psi^* \psi.
\label{Newton-phi}
\end{eqnarray}
\label{Newton-psiphi}%
\end{subequations}
If $V_\mathrm{eff} = 0$, these are the {\it  Schr\"odinger-Poisson equations}. If $V_\mathrm{eff} = (2 \pi a/m_a) (\psi^*\psi)^2$, where $a$ is the S-wave scattering length, Eqs.~\eqref{Newton-psiphi} are the {\it  Gross-Pitaevskii-Poisson (GPP) equations}. The energy is the sum of kinetic, self-interaction, and gravitational terms:
\begin{equation}
E  = 
 \int \!\!d^3r\, \left( \frac{1}{2m_a}  \nabla \psi^* \cdot \nabla \psi
+ V_\mathrm{eff}(\psi^* \psi)  +m_a\psi^*\psi \,  \Phi \right) .
\label{Egravity-psi}
\end{equation}
The total energy  in the relativistic theory can be approximated by the mass energy $N m_a$, where $N$ is the number of nonrelativistic axions in Eq.~\eqref{Naxion}, or, more accurately, by $Nm_a + E$. In the simplest spherically symmetric solutions, $\Phi(r)$ is time-independent and $\psi(r,t)$ has the harmonic time dependence in Eq.~\eqref{psi-periodic}. The asymptotic solution to Eq.~\eqref{Newton-phi} for the gravitational potential  as $r \to \infty$ is $\Phi(r) \to -GNm_a/r$, where $N$ is the axion number  in Eq.~\eqref{Naxion}.

The periodic spherically symmetric solutions of Eqs.~\eqref{Newton-psiphi} are uniquely determined by the central number density $n_0$, which can have any positive value. In some ranges of $n_0$, the solutions are unstable to small perturbations. Insight into stability under spherically symmetric perturbations can be obtained from Poincare's theory of linear series of equilibria  \cite{Katz:1978}. The number of unstable modes can only change at a critical point where  $d\varepsilon_b/dn_0$ is infinite. At such a critical point,  the number of unstable modes changes either by $+1$ or $-1$.

\section{SELF-BOUND SYSTEMS}
\label{sec:BoundAxions}

After a brief discussion of self-bound fluids, we discuss  {\it oscillons}, which are self-bound configurations in a real scalar field theory, and {\it axitons}, which are oscillons in a field theory for axions.

\subsection{Self-bound Fluids}

Many-body systems of particles that are self-bound fluids arise in various areas of physics. Atomic nuclei are quantum systems consisting of protons and neutrons bound by the nuclear force of QCD. They range from few-body clusters to liquid drops of nuclear matter consisting of hundreds of nucleons. The free expansion into the vacuum of a gas of helium-4 atoms produces bound systems ranging from few-atom helium molecules \cite{Schollkopf:1996} to droplets of superfluid helium \cite{Gomez:2011}. Self-bound droplets  of atoms bound by dipolar interactions have recently been produced in experiments with ultracold potassium-39 atoms \cite{Schmitt:2016,Semeghin:2018}.

A self-bound fluid can be described in a field theory by a classical solution that remains localized in space as time evolves. Since free field theories have no such classical solutions, the localized  solutions are bound by the self-interactions of the fields. A {\it soliton} is a localized classical solution of a field theory that has particle-like properties. The strict definition of a soliton requires that 
(1) it propagates unchanged except in its position,
(2) it emerges from collisions unchanged except perhaps for a time delay.
A more flexible definition might require only the first condition.

The classic example of a field theory with solitons is the sine-Gordon model in 1+1 space-time dimensions (1D). The soliton in its rest frame has a unique time-independent solution $\phi(x)$ whose total energy is $8f_a^2m_a$. The sine-Gordon model also has localized solutions $\phi(x,t)$ with larger energies called {\it breathers}, which in their rest frames are periodic functions of time with any angular frequency smaller than $m_a$.

Derrick's Theorem guarantees that  in any space-time dimension $d$ higher than 2, a field theory with a Lagrangian of the form in Eq.~\eqref{L-phi} or its generalizations with a multi-component field cannot have a soliton solution that is time-independent in its rest frame \cite{Derrick:1964}. It follows from a simple scaling argument that shows that the energy of any time-independent solution $\phi(\bm{r})$ can be lowered by decreasing its size. In particular, the rescaled solution $\phi(\lambda \bm{r})$ has gradient energy that scales like $\lambda^{2-d}$ and potential energy that scales like $\lambda^{-d}$. If $d>2$, the total energy can therefore be made arbitrarily small by increasing $\lambda$.

There are ways to evade Derrick's Theorem. The Skyrme model is a field theory  in 3+1 space-time dimensions (3D) whose field $U(x)$  takes its values in the 3-dimensional Lie group $SU(2)$ \cite{Skyrme:1962}. Its quanta can be interpreted as the pions of QCD. The Skyrme model has a topological quantum number whose integer values can be interpreted as baryon number. Its solitons with baryon number 1 are called {\it skyrmions}, and they can be identified with nucleons. It also has soliton solutions with larger and larger baryon numbers that can be identified with nuclei \cite{Braaten:1989rg}. The Skyrme model evades Derrick's Theorem by having a field $U(x)$ with values on a curved manifold.

\subsection{Oscillons}
\label{sec:Oscillon}

Bogolubsky and Makhankov discovered in 1976 that if the potential $V(\phi)$ in the Lagrangian in Eq.~\eqref{L-phi} has a suitable form, there are classical solutions that remain approximately localized and approximately periodic in time for a very large number of oscillation periods  \cite{Bogolubsky:1976}. They referred to such a solution as a  {\it pulson}. Pulsons were subsequently studied by Gleiser, who proposed the name {\it oscillon} \cite{Gleiser:1993pt}. They have also been called {\it quasibreathers} \cite{Fodor:2006zs}. Oscillons evade Derrick's theorem through their time dependence. Oscillons were discovered in the real scalar field theory with a symmetric double-well potential \cite{Bogolubsky:1976}. Oscillons were subsequently discovered in field theories with other potentials $V(\phi)$, including the sine-Gordon model \cite{Piette:1997hf}. A necessary condition for oscillons is that the potential satisfy $V''(\phi)<0$ for some region of $\phi$.

Oscillons have often been studied by starting from a spherically symmetric  initial configuration $\phi(r,0)$. The classical field equations are then solved numerically until the solution $\phi(r,t)$ can be approximated by a localized periodic configuration. The time evolution of the solution has three stages:
(1) the {\it relaxation stage}, in which it relaxes to an oscillon configuration by radiating away a significant fraction of its initial energy into outgoing waves,
(2) the {\it oscillon stage}, in which it remains stable for a very large number of oscillations, while slowly emitting outgoing waves,
(3) the {\it decay stage}, in which  it suddenly becomes unstable, and disappears into outgoing waves.
In the oscillon stage, the classical solution has properties that change slowly in time. The basic instantaneous properties of the oscillon include
\begin{itemize}
\item
the angular {\it oscillation frequency} $\omega$, 
\item
the {\it central energy density} $\rho_0$,
\item
the total energy or {\it mass} $M= \int \!d^3r\, \mathcal{H}$, where $\mathcal{H}$ is the Hamiltonian density in Eq.~\eqref{H-phi},
\item
the {\it radius $R_{99}$}  that encloses 99\% of the energy.
\end{itemize}
As time proceeds, the frequency $\omega$ steadily increases toward $m_a$ and $M$ steadily decreases.

Approximate solutions for oscillons can  be obtained by using an asymptotic expansion. The periodic solution for a spherically symmetric oscillon with angular frequency $\omega = m_a\sqrt{1-\epsilon^2}$ has an asymptotic expansion of the form \cite{Fodor:2008es}
\begin{equation}
\phi(r,t)=  \sum_{n=1}^\infty \epsilon^n \, \phi_n\big(\epsilon m_a r, \sqrt{1-\epsilon^2}\, m_a t\, \big),
\label{asymptoticexpansion}
\end{equation}
where each of the functions $\phi_n$ is localized. If the potential $V(\phi)$ is an even function of $\phi$, the sum is over odd powers $n$ only. The first function $\phi_1$ has the separable form of the  harmonic approximation in Eq.~\eqref{phi-harmonic}. The basic properties of the oscillon have asymptotic expansions in $\epsilon$ \cite{Fodor:2008es}. If $V(\phi)$ has the even power series in Eq.~\eqref{V-series}, the leading terms in the expansions are
\begin{subequations}
\begin{eqnarray}
\rho_{0} &=& (75.3\,  \epsilon^2/|\lambda_4|)~(m_a f_a)^2,
\\
M &=&  (75.6/|\lambda_4| \epsilon)~f_a^2/m_a, ~~~
\label{M-unstableoscillon}
\\
R_{99} &=& (2.75/\epsilon)/m_a.
\end{eqnarray}
\label{unstableoscillon}
\end{subequations}

An important issue is the lifetime of the oscillon before it disappears into outgoing waves. The oscillon lifetime has been studied in the field theory with a symmetric double-well potential  using an initial Gaussian configuration $\phi(r,0)$ with an adjustable radius $r_0$. Copeland et al.\  found that oscillons with lifetimes of over 1000 oscillations could be produced using any value of $r_0$ in a range that extended  by about a factor of 2 \cite{Copeland:1995fq}. Honda and Choptuik showed that there are critical values of  $r_0$ in this range near which the oscillon lifetime increases dramatically \cite{Honda:2001xg}. Near a critical radius $r_{0*}$, the lifetime appears to be linear in $\log(1/|r_0-r_{0*}|)$, which diverges as $r_0$ approaches $r_{0*}$. 

The spectrum of scalar waves emitted by the oscillon has been studied by Salmi and Hindmarsh in the model with a symmetric double-well potential \cite{Salmi:2012ta}. They found that the power spectrum of radiation with angular frequency $\omega$ from an oscillon with  frequency $\omega_0$ just below $m_a$ has sharp  peaks near integer harmonics of $\omega_0$. The largest peak is near $2 \omega_0$. The power in each successive higher harmonic is smaller by orders of magnitude. The second strongest peak in $\omega$ is near $m_a$. They explained this peak by the oscillation frequency of the oscillon having a distribution that can be modeled by a Breit-Wigner resonance centered at $\omega_0$  with a tail that extends above $m_a$. If $V(\phi)$ is an even function of $\phi$ with respect to its minimum, as in the sine-Gordon model, the peaks are at odd-integer harmonics of $\omega_0$ \cite{Salmi:2012ta}. Peaks in the radiation spectrum at odd-integer harmonics of $\omega_0$ have also been observed in a model whose potential $V(\phi)$ has an attractive $\phi^4$ term and a repulsive $\phi^6$ term \cite{Mukaida:2016hwd}. 

If the  oscillon configuration has only long wavelengths much larger than $2\pi/m_a$ and if the frequency difference $m_a - \omega$ is small compared to $m_a$, the oscillon can be described more simply by the complex field $\psi(\bm{r},t)$ of NREFT. If we omit terms in the Hamiltonian that are suppressed  by additional gradients of $\psi$, the classical field equation reduces to Eq.~\eqref{psi-EOM}. This equation has time-periodic solutions with the form in Eq.~\eqref{psi-periodic}. The basic properties of the oscillon in NREFT include
\begin{itemize}
\item
the {\it binding energy} $\varepsilon_b$ of a boson, which corresponds to the oscillation frequency $\omega =m_a - \varepsilon_b$,
\item
the {\it central number density}  $n_0$,
\item
the {\it particle number} $N$ in Eq.~\eqref{Naxion},
\item
the {\it total binding energy} $E_b = - \int\!d^3r\, \mathcal{H}_\mathrm{eff}$, 
\item
the radius $R_{99}$  that encloses 99\% of the  particle number.
\end{itemize}
The total central energy density $\rho_0$ of the oscillon can be approximated by the central mass density $m_a n_0$. The mass $M$ of the oscillon can be approximated by  $N m_a$ or, more accurately, by $ N m_a-E_b$. 

\subsection{Axitons}
\label{sec:Axiton}

The axion field theory with the instanton potential in Eq.~\eqref{V-instanton} is just the sine-Gordon model in  3D. Since the chiral potential in Eq.~\eqref{V-chiral} has the same qualitative behavior as the instanton potential, the axion field theory with the chiral potential also has oscillon solutions. An oscillon in an axion field theory can also be called an {\it axiton}. Naive approximations to the spherically symmetric oscillon solutions $\phi(r,t)$ have been obtained using the harmonic approximation in Eq.~\eqref{phi-harmonic} \cite{Visinelli:2017ooc}. Accurate solutions could be obtained by solving the infinite set of coupled equations for the harmonics $\phi_{2n+1}(r)$ in Eq.~\eqref{phi-cosine} \cite{Piette:1997hf} or by solving Eq.~\eqref {phi-EOM} for the time-dependent field $\phi(r,t)$ directly.

\begin{figure}[t]
\centerline{ \includegraphics*[width=8cm,clip=true]{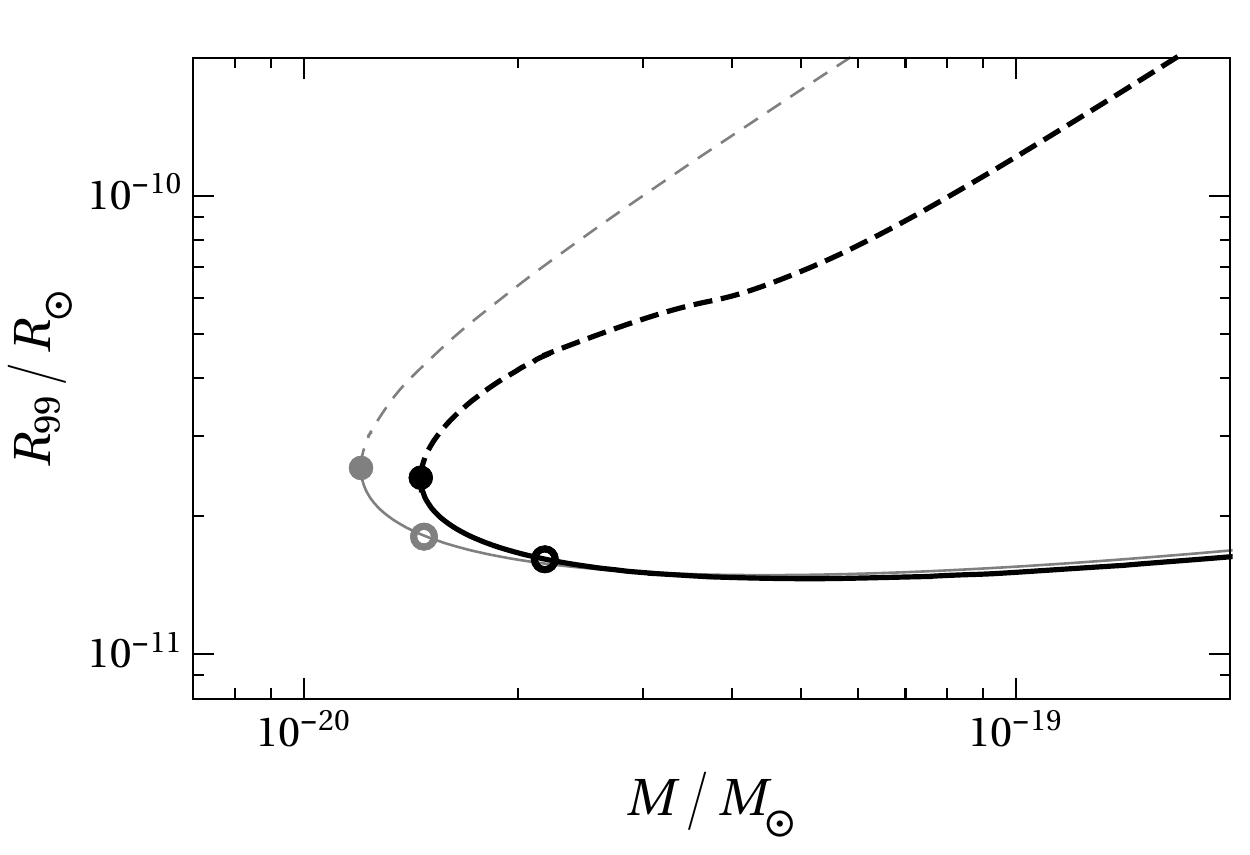} }
\vspace*{0.0cm}
\caption{Radius $R_{99}$ versus the mass $M= N m_a$ for the oscillon of the axion field. The axion mass is $m_a = 10^{-4}$~eV. The axion potential is either the chiral potential with $z=0.48$ (black lines) or the instanton potential (gray lines). The solid dot is the critical point that separates the unstable branch (dashed line) from the locally stable  branch (solid line). The open dot marks the point to the left of which the lower branch is unstable to large fluctuations.}
\label{fig:RvsM-axiton}
\end{figure}

An  oscillon configuration in which gradients are always small compared to $m_a$ can be described more simply using the complex field $\psi(\bm{r},t)$ of NREFT. The simple equation of motion in Eq.~\eqref{psi-EOM} has spherically symmetric solutions $\psi(r,t)$ with the harmonic time dependence in Eq.~\eqref{psi-periodic}. This equation for $\psi(r,t)$ with the naive effective potential $V_\mathrm{eff}$ in Eq.~\eqref{V-naive} is equivalent to the harmonic approximation for $\phi(r,t)$. The numerical results for oscillons presented below are calculated by solving this equation using the naive chiral effective potential in Eq.~\eqref{Veff-chiral} with $z=0.48$ or using the naive instanton effective potential in Eq.~\eqref{Veff-instanton}.

In Fig.~\ref{fig:RvsM-axiton}, we show the radius $R_{99}$ versus the mass $M$ as the central number density $n_0$ is increased. There are two branches of solutions that meet at a critical point indicated by the solid dot. As $n_0$ increases, the solutions approach the critical point along the upper unstable branch and then move away from the critical point along the locally stable lower branch. In the region well before the critical point where $R_{99}$ versus $M$ is a straight line on a log-log plot, the basic properties of the unstable oscillon are given by the leading terms in the asymptotic expansions in Eqs.~\eqref{unstableoscillon}. For the chiral potential with $z=0.48$, the basic properties of the oscillon at the critical point  are
\begin{subequations}
\begin{eqnarray}
&& \varepsilon_{b*} = 0.050~m_a, \qquad ~~~
n_{0*} = 13.9~m_a f_a^2,
\\
&&E_{b*} =  -26.8~f_a^2/m_a, ~~~
N_* =  488~f_a^2/m_a^2,
\\
&&R_{99*} = 8.6/m_a.
\end{eqnarray}
\label{criticalaxiton}%
\end{subequations}
(For the instanton potential, the five numerical coefficients above are 0.052, 6.5, $-8.9$, 402, and 9.0.) 
For the chiral potential with $z=0.48$ and $m_a = 10^{-4\pm 1}$~eV, the critical number of axions is 
$N_* = 2 \times 10^{50\mp 4}$. The critical mass $N_* m_a$  is  $3 \times 10^{10\mp 3}$~kg, 
and the critical radius $R_{99*}$ is $2 \times 10^{-2\mp 1}$~m.

The total binding energy $E_b$ of the oscillon is negative at the critical point. The negative sign indicates that the oscillon is unstable to sufficiently large fluctuations: there are configurations of outgoing waves with the same number $N$ of bosons and lower total energy. The total binding energy changes sign to positive at a point indicated in Fig.~\ref{fig:RvsM-axiton} by the open dot. Beyond the open dot, the oscillon is stable even to large fluctuations. For the chiral potential with $z=0.48$, the basic properties of the oscillon with zero binding energy  are $\varepsilon_{b} = 0.144\,m_a$, $n_{0} = 22.1\,m_a f_a^2$, $N =  729\,f_a^2/m_a^2$, and $R_{99} = 5.7/m_a$. (For the instanton potential, the four numerical coefficients above are 0.116, 13.0, 494, and 6.4.) 

If the mass $M$ of the oscillon is much larger than the critical value $M_*$, there is  a scaling region in which its basic properties scale as powers of $M$. For the chiral potential with $z=0.48$, the basic properties of the oscillon in the scaling region  are
\begin{subequations}
\begin{eqnarray}
 \varepsilon_{b} &=& 0.5\, m_a,
\label{epsilonb-scaling} 
\\
n_{0} &\approx& 25~(M/M_*)^{1.5} m_a f_a^2,
\\
E_{b} &=&  240~(M/M_*)^{2.7} f_a^2/m_a,
\\
R_{99} &=& 2.8~(M/M_*)^{1.2}/m_a.
\end{eqnarray}
\label{axiton-scaling}%
\end{subequations}
(For the instanton potential, the four numerical coefficients above are  0.5, 18, 200, and 2.8.) The central density $n_0$ has much larger fluctuations around the scaling behavior than the other properties. The size of the boson binding energy $ \varepsilon_{b}$ in Eq.~\eqref{epsilonb-scaling} makes the accuracy of the nonrelativistic approximation questionable.

NREFT provides a simple explanation for a puzzling feature of oscillons: their rather sudden decay into outgoing waves. The mass of an oscillon decreases steadily because of the emission of scalar waves with relativistic frequencies. An oscillon on the stable branch in Fig.~\ref{fig:RvsM-axiton} therefore moves steadily to the left. When it reaches the point where the binding energy is 0, it becomes unstable to large fluctuations and it can  disappear into outgoing waves with nonrelativistic  frequencies near $m_a$. If the oscillon reaches the critical point, it becomes unstable to small fluctutions and it must disappear into outgoing nonrelativistic waves. The explanation for the sudden decay of the oscillon in terms of the real scalar field is more complicated \cite{Gleiser:2008ty}.

Studies of the time evolution of the axion field in the expanding universe by Kolb and Tkachev revealed the existence of localized solutions with the three stages described in Section~\ref{sec:axionMC} \cite{Kolb:1993zz,Kolb:1993hw}. The second stage consists of multiple cycles of growth in the amplitude of $\phi$ to a value near $\pi f_a$ followed by sudden collapse. Kolb and Tkachev used the word {\it axiton} to refer to these 3-stage localized solutions. We suggest that a more useful definition  of an axiton is an oscillon in an axion field theory. By this definition, axiton would refer to the axion field during the growth phase of one of the multiple cycles. The sudden collapse of the axiton is caused by its central density decreasing to below the critical value because of the expansion of the universe.

\section{GRAVITATIONALLY BOUND SYSTEMS}
\label{sec:GravityBound}

We first describe the simplest {\it boson stars}, which are gravitationally bound BEC's of bosons that have no self-interactions. We then discuss {\it axion stars}, which are bound BEC's of axions with gravity.

\subsection{Boson Stars}
\label{sec:Bosestar}

A {\it boson star} is a gravitationally bound system of bosons. Reviews of boson stars from a general relativity perspective have been presented in Refs.~\cite{Jetzer:1991jr,Liebling:2012fv}. The simplest boson stars consist of identical bosons  whose number is conserved. Such a boson star can be described by a classical solution of the  {\it Einstein-Klein-Gordon equations} with a complex field $\Psi(x)$. These equations have black-hole solutions with $\Psi=0$ whose size is roughly the Schwarzchild radius $2GM$. They also have boson-star solutions with much larger radius that are localized and spherically symmetric, with a time-independent the metric tensor and a harmonic scalar field: $\Psi(\bm{r},t) = \Psi_0(r) e^{-i \omega t}$. There is a critical mass $M_*$ of the boson star above which there are no such solutions. A boson star with mass greater than $M_*$ will either collapse into a black hole or decrease its mass to below $M_*$ by scalar field radiation. The solutions for boson stars were first studied in Refs.~\cite{Kaup:1968zz,Ruffini:1969qy}. The critical mass was first determined accurately by Breit, Gupta, and Zaks: $M_* = 0.633/(Gm_a)$ \cite{Breit:1983nr}. If the boson mass is $m_a= 10^{-4}$~eV, the critical mass is $8 \times 10^{-7}~M_\odot$, where $M_\odot$ is the mass of the sun. 

Identical bosons whose number is not conserved can also be bound gravitationally into a boson star. The boson star can be described by a classical solution of the {\it Einstein-Klein-Gordon equations} in Eqs.~\eqref{EOM-gravity} for a real scalar field $\phi(x)$ with $V' = m_a^2 \phi$. There are black-hole solutions with $\phi=0$. All periodic solutions with nonzero $\phi$ have infinite energy, because they must have standing scalar waves extending to infinity, with the flux of the incoming waves balancing that of the outgoing waves. Seidel and Suen discovered in 1991 that there are boson-star solutions that can be accurately approximated by localized periodic solutions \cite{Seidel:1991zh}. They referred to such solutions that are approximately periodic and approximately localized as {\it oscillatons}. The critical mass for a boson star with a non-self-interacting real scalar field is $M_* = 0.607/(Gm_a)$   \cite{UrenaLopez:2002gx}. This is about 4\% smaller than the critical mass of a boson star with a complex scalar field.

If the mass $M$ of the boson star is much smaller than $M_*$, the bosons are all nonrelativistic. In this limit, the equations for a boson star with either a real scalar field or a complex scalar field can be reduced to the Schr\"odinger-Poisson equations given by Eqs.~\eqref{Newton-psiphi} with $V_\mathrm{eff} = 0$. The boson star is a gravitationally bound BEC in which all the bosons are in the same quantum state with the wavefunction $\psi(\bm{r},t)$. Chavanis and Delfini obtained simple scaling results for some basic properties of the boson star as functions of its mass $M = N m_a$  in the limit $G M m_a \ll 1$ \cite{Chavanis:2011zm}. The basic properties of the nonrelativistic boson star consisting of bosons with no self-interactions are
\begin{subequations}
\begin{eqnarray}
\varepsilon_b &=& 0.159~(G M m_a)^2 m_a,
\label{epsb-lowmass}
\\
\rho_0 &=& 4.23\times 10^{-3}~ (G M m_a)^3 M m_a^3,
\label{rho0-lowmass}
\\
E_b &=& 6.15~ (G M m_a)^2 M,
\label{Eb-lowmass}
\\
R_{99} &=& 10.05~ (G M m_a)^{-1}/m_a.
\label{R99-lowmass}
\end{eqnarray}
\label{axionstar-lowmass}%
\end{subequations}
As $M$ increases, the radius $R_{99}$ decreases, which is somewhat counterintuitive.

\subsection{Dilute Axion Stars}
\label{sec:Dilute}

An {\it axion star} is a boson star that consists of axions, a possibility first considered by Tkachev \cite{Tkachev:1991ka}. The potential energy provided by axion self-interactions can alter the balance of forces in the star. The interaction between a pair of axions gives an attractive force, because the scattering length $a$ in Eq.~\eqref{a-fa} is negative. The classical solutions for an axion star  can be determined by solving the Einstein-Klein-Gordon equations in Eq.~\eqref{EOM-gravity} for a real scalar  field $\phi(\bm{r},t)$ with axion potential $V(\phi)$. The solutions are approximately  localized  and approximately periodic, so they can be called oscillatons. Solutions for spherically symmetric axion stars  were first calculated numerically by Barranco and Bernal \cite{Barranco:2010ib}. They found solutions only below a  critical mass $M_*$. We refer to the  solutions as {\it dilute axions stars}, because the energy density of axions is much less than the QCD scale $(m_a f_a)^2$, even at the center of the star.

The equations for dilute axion stars can be greatly simplified without much loss of accuracy. The axions are nonrelativistic, so they can be described by the complex scalar field $\psi(\bm{r},t)$ of NREFT. The number density $\psi^*\psi$ is small compared to $m_a f_a^2$, so the effective potential can be approximated by the leading term in its power series:
\begin{equation}
V_\mathrm{eff}(\psi^* \psi)  \approx \frac{2 \pi  a}{m_a}( \psi^* \psi)^2
= \frac{\lambda_4 }{16f_a^2}( \psi^* \psi)^2.
\label{Veff-approx}
\end{equation}
Finally, the gravitational forces are weak, and they can be described by Newtonian gravity. Dilute axion stars can therefore be described accurately by the GPP equations in Eqs.~\eqref{Newton-psiphi}. These equations were used by Chavanis to derive simple variational approximations to basic properties of dilute axion stars \cite{Chavanis:2011zi}.

\begin{figure}[t]
\centerline{ \includegraphics*[width=8cm,clip=true]{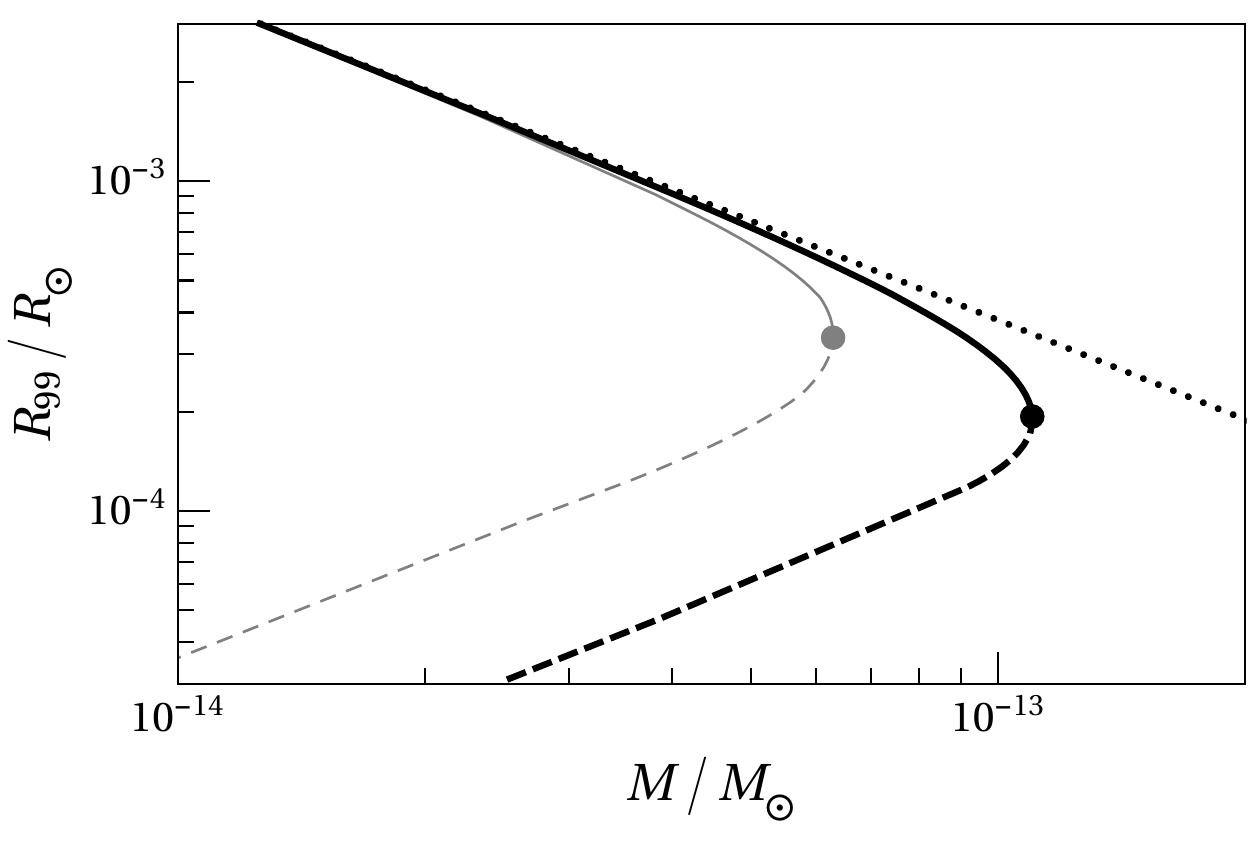} }
\vspace*{0.0cm}
\caption{Radius $R_{99}$ versus mass $M$ for a dilute axion star. The axion mass is $m_a = 10^{-4}$~eV. The axion potential is the chiral potential with $z=0.48$ (black lines) or the instanton potential (gray lines). The dot is a critical point that separates the stable branch (solid line) from the unstable branch (dashed line). The dotted line is for a boson star with no self-interactions.}
\label{fig:RvsM-dilute}
\end{figure}

The dependence of the radius $R_{99}$ of the dilute axion star on the mass $M$ is illustrated in Fig.~\ref{fig:RvsM-dilute}. There are two branches of solutions  that meet at the critical point indicated by the solid dot. As the central number density $n_0$ increases, the solutions approach the critical point along the stable upper branch and then move  away from the critical point along the unstable lower branch. The number of spherically symmetric unstable modes changes from 0 to 1 at the critical point. Well before the critical point, the basic properties of the stable dilute axion star reduce to those in Eq.~\eqref{axionstar-lowmass} for a  boson star consisting of  nonrelativistic bosons with no self-interactions.

The GPP equations were used by Chavanis and Delfini to derive scaling approximations to basic properties of a dilute axion star at the critical mass  \cite{Chavanis:2011zm}. The basic properties of the critical dilute axion star are 
\begin{subequations}
\begin{eqnarray}
\varepsilon_{b*} &=& \big(36/| \lambda_4| \big) \big(Gf_a^2 \big)  m_a,
\label{epsb-critical}
\\
n_{0*} &=& \big(404/\lambda_4^2 \big) \big(Gf_a^2 \big) m_a f_a^2,
\label{rh0-critical}
\\
E_{b*}  &=& \big(90.7/|\lambda_4|^{3/2} \big) \big(Gf_a^2 \big)^{1/2} f_a^2/m_a,
\label{Eb-critical}
\\
N_* &=& \big(10.15/|\lambda_4|^{1/2}\big) \big(Gf_a^2 \big)^{-1/2} f_a^2/ m_a^2,
\label{M-critical}
\\
R_{99*} &=& \big(0.55\, |\lambda_4|^{1/2} \big) \big(Gf_a^2 \big)^{-1/2}/ m_a.
\label{R99-critical}
\end{eqnarray}
\label{axionstar-critical}%
\end{subequations}
If the axion mass is $m_a = 10^{-4\pm 1}$~eV, the critical number with the chiral potential is $N_* =1.2 \times 10^{57\mp 3}$. (For the instanton potential, the critical number is smaller by the factor 0.59 because of the different value of $\lambda_4$.) The critical mass $N_* m_a$  with the chiral potential  is $1.1 \times 10^{-13\mp 4}\,M_\odot$, where $M_\odot$ is the mass of the sun. The critical radius $R_{99*}$ is $1.9 \times 10^{-4}R_\odot$, where $R_\odot$ is the radius of the sun.

\begin{figure}[t]
\centerline{ \includegraphics*[width=8cm,clip=true]{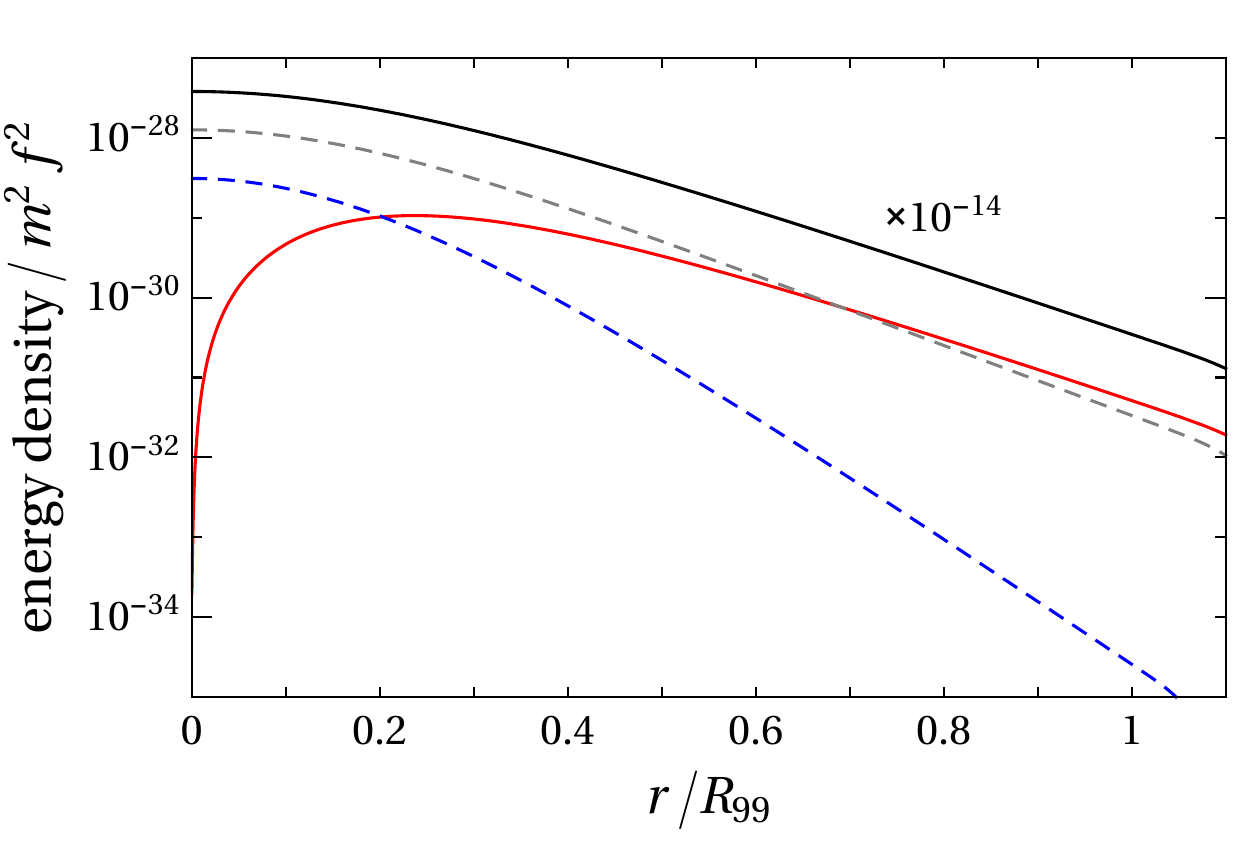} }
\vspace*{0.0cm}
\caption{
Contributions to the energy density $\rho(r)$ in the critical dilute axion star as functions of the radial coordinate $r$. Their absolute values are shown on a log scale, with positive (negative) contributions  shown as solid (dashed) curves. The axion potential is the chiral potential with $z=0.48$ and $m_a = 10^{-4}$~eV. The curves in order of decreasing size at small $r$ are mass (black, multiplied by $10^{-14}$), gravitational (gray), potential (blue), and kinetic (red).}
\label{fig:rhovsr-dilute}
\end{figure}

The contributions to the energy density of the critical dilute axion star are shown as functions of the radial coordinate $r$  in Fig.~\ref{fig:rhovsr-dilute}. The mass density $m_a \psi^*\psi$ is many orders of magnitude larger than the other contributions to the energy density. The gravitational and potential energy densities are comparable at small $r$. The kinetic and gravitational energy densities are comparable in absolute value at large $r$. Thus the balance of forces in the critical dilute axion star is between the  attractive forces from gravity and from axion self-interactions and the repulsive force from the kinetic pressure of the axions.

\begin{figure}[t]
\centerline{ \includegraphics*[width=8cm,clip=true]{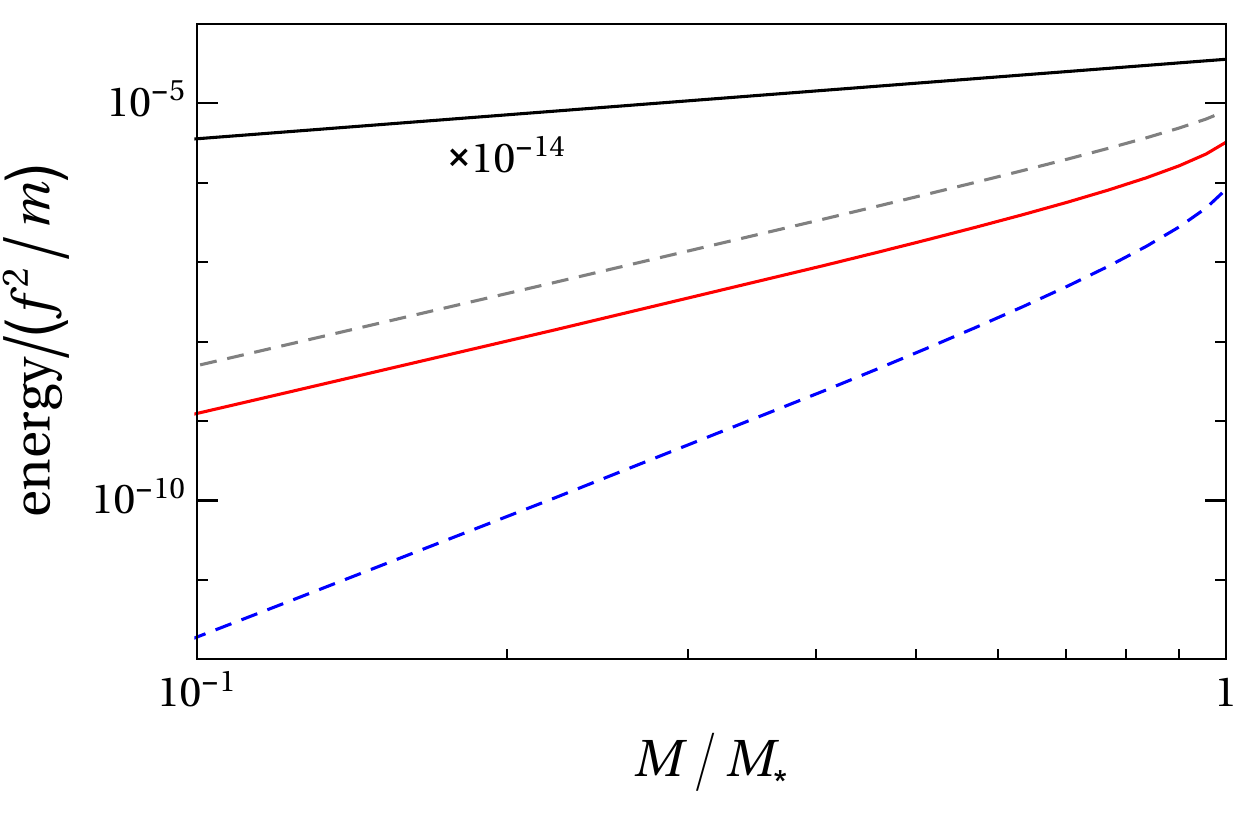} }
\vspace*{0.0cm}
\caption{
Contributions to the total energy of the stable dilute axion star as functions of its mass $M$. Their absolute values are shown on a log scale, with positive (negative) contributions  shown as solid (dashed) curves. The curves in order of decreasing size are mass (black, multiplied by $10^{-14}$), gravitational (gray), kinetic (red), and potential (blue).}
\label{fig:Energy-nhat:dilute}
\end{figure}

The contributions to the total energy of the dilute axion star are shown in Fig.~\ref{fig:Energy-nhat:dilute} as functions of its mass $M$. The mass energy $N m_a$ is many orders of magnitude larger than the other contributions. The potential energy becomes increasingly important as $M$ approaches the critical mass $M_*$.

\subsection{Dense Axion Stars}
\label{sec:Dense}

\begin{figure}[t]
\centerline{ \includegraphics*[width=8cm,clip=true]{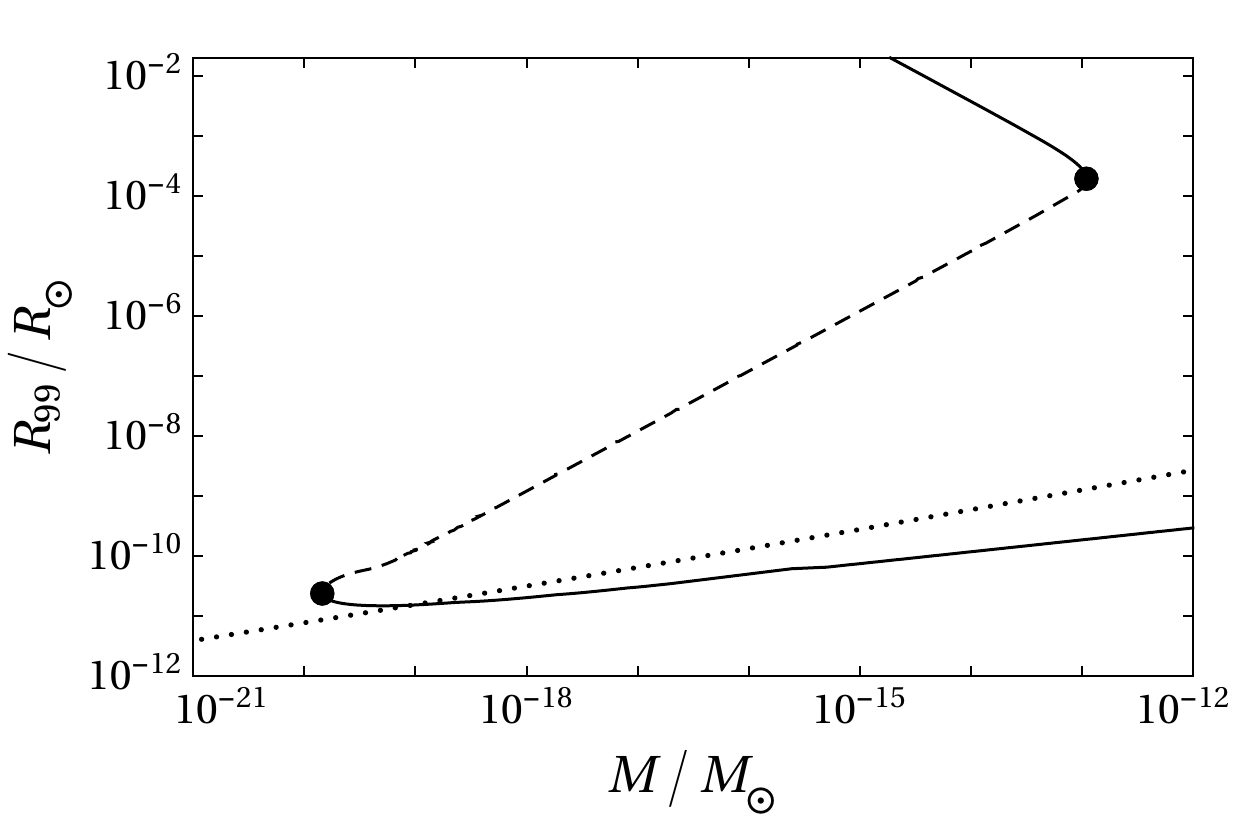} }
\vspace*{0.0cm}
\caption{Radius $R_{99}$ versus mass $M$ for an axion star. The axion potential is the chiral potential with $z=0.48$ and $m_a = 10^{-4}$~eV. The dots are critical points that separate the stable branches (solid lines) from the unstable branch (dashed line). The dotted line is the Thomas-Fermi approximation.}
\label{fig:RvsM-both}
\end{figure}

It was pointed out in Ref.~\cite{Braaten:2015eeu}  that there could be other branches of axion star solutions in addition to the dilute axion stars. Evidence for another branch was found by following the unstable solution that emerges from the critical point for the dilute axion star in Fig.~\ref{fig:RvsM-dilute} as the central density $n_0$ is increased further. The field equations solved in Ref.~\cite{Braaten:2015eeu} were those for the complex field $\psi(r)$ and the gravitational potential $\Phi(r)$  in Eqs.~\eqref{Newton-psiphi} with the naive effective instanton potential in Eq.~\eqref{Veff-instanton}. The mass $N m_a$ and the radius $R_{99}$ were calculated as functions of the central number density $n_0$. The corresponding results using the naive effective chiral potential in Eq.~\eqref{Veff-chiral} are shown in Fig.~\ref{fig:RvsM-both}. After the radius decreases by about 7 orders of magnitude, there is a second critical point where $d\varepsilon_b/dn_0$ is infinite. By Poincare's theory of linear series of equilibria \cite{Katz:1978}, the number of unstable modes must change from 1 to either 2 or 0 at the second critical point. The solutions near and beyond the second critical point were called  {\it dense axion stars} in Ref.~\cite{Braaten:2015eeu}, because the mass density  $m_a \psi^*\psi$ inside the axion star becomes comparable to the QCD scale $(m_a f_a)^2$. The effective potential $V_\mathrm{eff}$ therefore cannot be truncated after the $(\psi^* \psi)^2$ term as in Eq.~\eqref{Veff-approx}.

The plot of $R_{99}$ versus $M$ for the dense axion star near the second critical point in Fig.~\ref{fig:RvsM-both} is almost indistinguishable from that for the oscillon in Fig.~\ref{fig:RvsM-axiton}. The reason for this is that the effects of gravity are almost negligible, as pointed out in Ref.~\cite{Visinelli:2017ooc}. Thus a dense axion star near the second critical point is actually an oscillon! The number of unstable modes changes from  1 to 0 at the second critical point. The basic properties of the critical dense axion star are the same as those for the critical oscillon in Eqs.~\eqref{criticalaxiton}.

\begin{figure}[t]
\centerline{ \includegraphics*[width=8cm,clip=true]{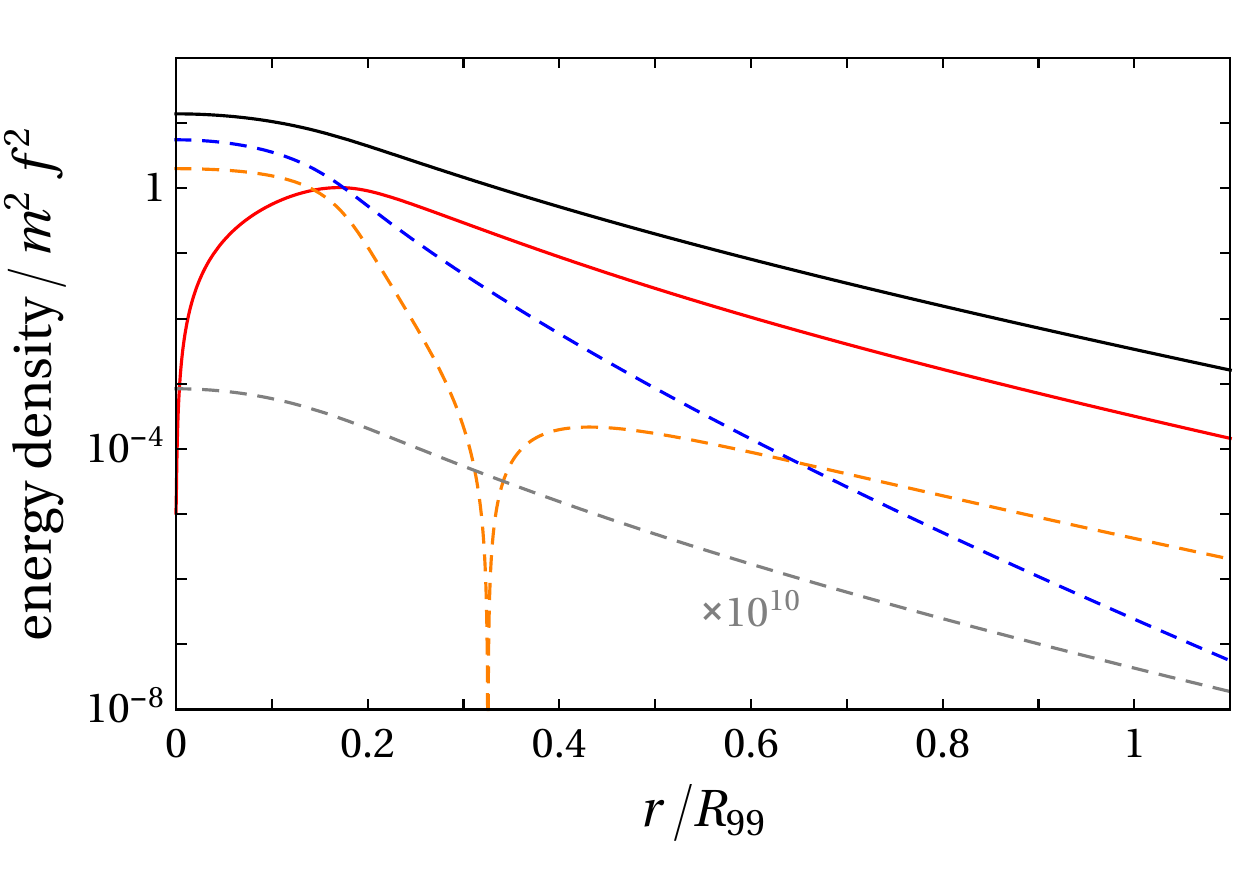} }
\vspace*{0.0cm}
\caption{
Contributions to the energy density $\rho(r)$ in the critical  dense axion star as functions of the radial coordinate $r$. Their absolute values are shown on a log scale, with positive (negative) contributions  shown as solid (dashed) curves. The curves in order of decreasing size at  $r=0.4\, R_{99*}$ are mass (black), kinetic (red), potential (blue), relativistic correction to kinetic (orange), and gravitational (gray, multiplied by $10^{10}$).}
\label{fig:Energy-r:axiton}
\end{figure}

The contributions to the energy density of the critical  dense axion star  or critical oscillon are shown as functions of the radial coordinate $r$  in Fig.~\ref{fig:Energy-r:axiton}. The largest contribution to the energy density at all $r$ is the mass density $m_a \psi^*\psi$. It is always larger than the kinetic energy density by at least a factor of 4.6. The second largest contribution is from the potential energy at small $r$ and from the  kinetic  energy  at large $r$. The gravitational contribution is smaller by many orders of magnitude, as first pointed out in Ref.~\cite{Visinelli:2017ooc}. The balance of forces is between the repulsive kinetic pressure and an attractive force from the axion effective potential.

Beyond the second critical point, the radius $R_{99}$ of the dense axion star begins to increase as a function of the mass $M$, as shown in Fig.~\ref{fig:RvsM-both}. As the central density continues to increase, the results from Eqs.~\eqref{Newton-psiphi} come close to matching on smoothly to the Thomas-Fermi approximation  \cite{Wang:2001wq}, which is the straight dotted line in Fig.~\ref{fig:RvsM-both}. In this approximation, the kinetic pressure is neglected except near the surface. In the interior, the attractive force from gravity is balanced instead by a repulsive force from the axion  effective potential. In Ref.~\cite{Braaten:2015eeu}, the Thomas-Fermi approximation was used to extrapolate the curve of $R_{99}$ to very large values of $M$.  As pointed out in Ref.~\cite{Visinelli:2017ooc}, the curve for $R_{99}$  versus $M$ actually crosses the line for the Thomas-Fermi approximation at a small angle. Thus the Thomas-Fermi approximation is not appropriate for dense axion stars.

\begin{figure}[t]
\centerline{ \includegraphics*[width=8cm,clip=true]{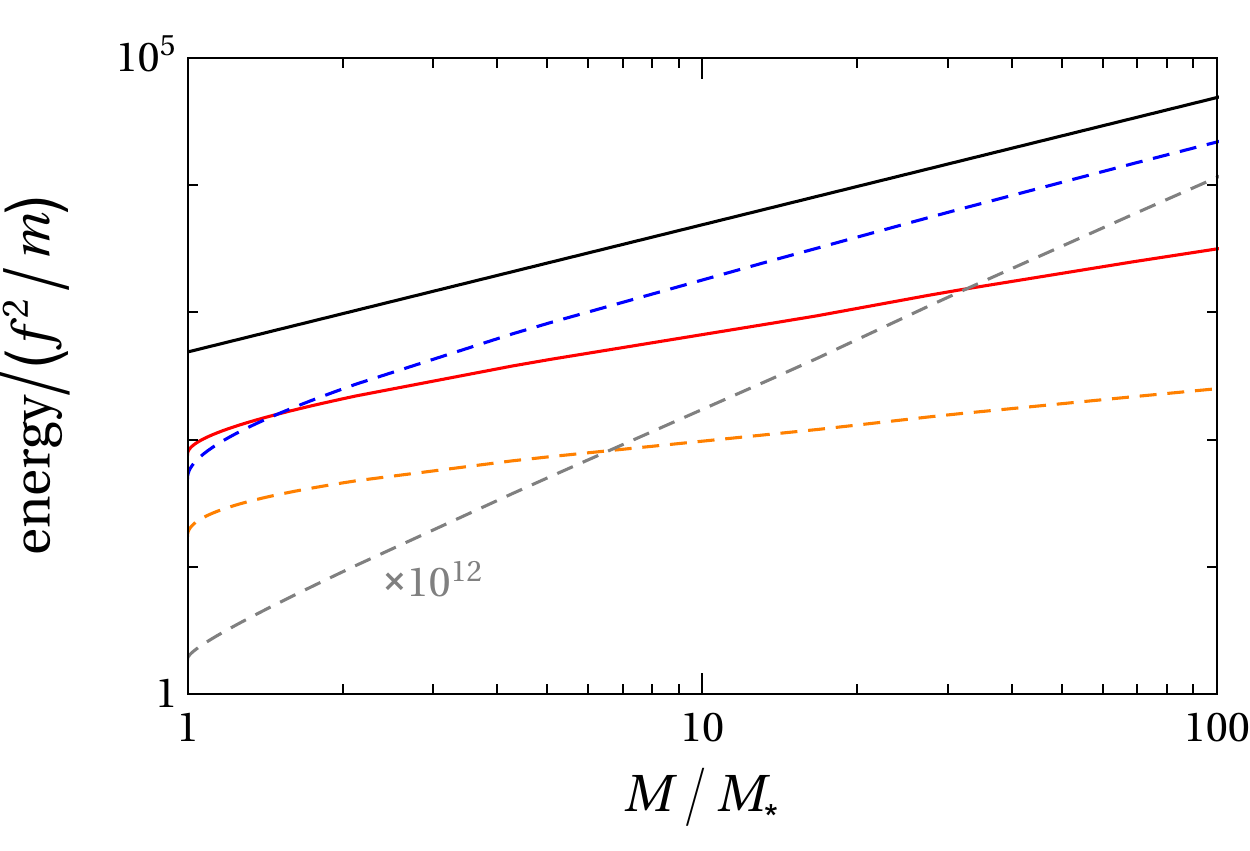} }
\vspace*{0.0cm}
\caption{
Contributions to the total energy of the locally stable dense axion star as functions of its mass $M = N m_a$. Their absolute values are shown on a log scale, with positive (negative) contributions  shown as solid (dashed) curves. The curves in order of decreasing size at $M = 3\, M_*$ are mass (black), potential (blue), kinetic (red), relativistic correction to kinetic (orange), and gravitational (gray, multiplied by $10^{12}$).}
\label{fig:Energy-nhat:axiton}
\end{figure}

The contributions to the total energy of the dense axion star are shown as functions of the mass $M$   in Fig.~\ref{fig:Energy-nhat:axiton}. As $M$ increases from the critical value $M_*$, the negative total binding energy $E_b$ increases. It reaches 0 when the mass is about $1.5\,M_*$, which is near where the potential and kinetic energies cross in Fig.~\ref{fig:Energy-nhat:axiton}. At larger $M$, the dense axion star is stable to large fluctuations as well as to small fluctuations. In the scaling region of $M$, the basic properties of the dense axion star are given in Eq.~\eqref{axiton-scaling}.

A simple picture for the fate of an isolated dense axion star emerges. A dense axion star corresponds to a point on the lower branch  in Fig.~\ref{fig:RvsM-both}. As time proceeds, the point moves to the left, because the axion star emits relativistic axions with momenta of order $m_a$. When it reaches the open dot in Fig.~\ref{fig:RvsM-axiton} where the total binding energy is 0, the axion star  becomes unstable to large fluctuations, and it can disappear into nonrelativistic axions with momenta much smaller than $m_a$. If it reaches the critical point, the axion star becomes unstable to small fluctuations and it must disappear into nonrelativistic axions. The lifetime of a dense axion star may be too short for it to be cosmologically significant as an astrophysical object. However dense axion stars can still have an important cosmological effect by transforming nonrelativistic axions into relativistic axions.

The gravitational contribution to the energy in Fig.~\ref{fig:Energy-nhat:axiton} is smaller than the other contributions by many orders of magnitude. Extrapolation of the curves to much larger values of $M$ implies that the gravitational energy may be comparable to the kinetic energy when the mass is about $10^{13}\, M_*$, which corresponds to about $10^{-7}\, M_\odot$. At much smaller masses, the oscillon description of the dense axion star should be accurate.

\section{ THEORETICAL ISSUES}
\label{sec:Issues}

In this section, we discuss theoretical issues involving isolated axion stars that have not yet been completely understood.

\subsection{Emission of Relativistic Axions from Axion Stars}
\label{sec:Emission}

Since the axion field is a real Lorentz scalar, the number of axions is not conserved. There are scattering reactions that change the number of axions and that can transform nonrelativistic axions into relativistic axions. A localized axion configuration with nonrelativistic wavelengths, such as an axion star, inevitably radiates axion waves with relativistic wavelengths. A bound configuration of axions therefore has a finite lifetime. It is important to understand the lifetime of  axion stars, because it  determines whether they can have any significance as astronomical objects.

NREFT  appears to give  unambiguous predictions for the loss rate of nonrelativistic axions from axion stars \cite{Braaten:2016dlp}. The rate of decrease in the number $N$ of nonrelativistic axions is determined by the anti-Hermitian terms in the effective Hamiltonian, and is given in Eq.~\eqref{dN/dt-X}. In a dilute axion star, the loss of axions is dominated by their decays into two photons. The decay rate of the dilute axion star is the same as the decay rate $\Gamma_a$ of the axion, which is given in Eq.~\eqref{Gamma}. The lifetime of a dilute axion star is therefore tens of orders of magnitude longer than the age of the universe. The lifetime of the dense axion star can be defined as the time required for its mass to decrease by a factor $1/e$ as it moves to the left along the lower branch in Fig.~\ref{fig:RvsM-both}. In a dense axion star, the loss  rate from inelastic axion reactions is much larger than that from decay into two photons. The contribution to $dN/dt$ from the $4 \to 2$ process is given by the second term in Eq.~\eqref{dN/dt-X}. This term is about 5 orders of magnitude larger than the first term. The resulting prediction for the lifetime of the dense axion star is still much longer than the age of the universe \cite{Braaten:2016dlp}.

The predictions of NREFT for the loss rate of nonrelativistic axions are incomplete. There are loss processes for axions in the relativistic theory that do not seem to be reproduced by NREFT. Gravity is an inessential complication for these loss processes, so we will discuss them in terms of oscillons. NREFT should correctly reproduce results from the relativistic theory for an oscillon with a small boson binding energy $\varepsilon_b \ll m_a$ as an expansion in powers of $\varepsilon_b/m_a$. However such an expansion is blind to  terms that are exponentially small in $m_a/\varepsilon_b$, such as $\exp(-c\sqrt{m_b/\varepsilon_b}\,)$, where $c$ is a constant. Thus we should not expect a loss rate having such an exponential factor to be reproduced by NREFT.

The existence of loss processes whose rates have exponentially small factors can be inferred from the asymptotic expansion for the oscillon in Eq.~\eqref{asymptoticexpansion}. The asymptotic expansion differs from the exact periodic solution by terms that are exponentially small in $\epsilon$ \cite{Segur:1987mg}. These terms are not localized: they have a radiative tail in the form of a standing wave with an exponentially small amplitude that extends to infinity and has infinite energy. In the absence of incoming waves, the outgoing waves decrease the total energy $M$ of the localized part of the solution. The rate $dM/dt$ of decrease in the mass of the oscillon with angular frequency $\omega = \sqrt{1-\epsilon^2} \, m_a$ in the limit $\epsilon \to 0$ has the form \cite{Fodor:2009kf}
\begin{equation}
-\frac{dM}{dt} = \frac{A}{\epsilon^2} \, \exp(-3.406/\epsilon) \, f_a^2,
\label{dE/dt-epsilon}
\end{equation}
where the prefactor $A$ depends on the axion potential $V(\phi)$. The sine-Gordon model is a special case in which $A$ is suppressed by $\epsilon^2$ \cite{Fodor:2009kf}. Fodor et al.\  determined $A$ for the sine-Gordon model in 3D: $A =760.5\,  \epsilon^2$. 

Eby et al.\ have derived an expression for the axion loss rate that can be expressed in terms of the complex field $\psi(x)$ of NREFT \cite{Eby:2015hyx}. Their derivation involves the matrix  element of $V(\phi)$ between an initial state of $N$ condensed axions, each with energy $\omega= m_a - \varepsilon_b$, and a final state consisting of $N-3$ condensed axions plus an on-shell axion with energy $3 \omega$. This can be interpreted as a $3 \to 1$ reaction, which is forbidden in the vacuum by conservation of energy and momentum. Their result for the rate of energy loss \cite{Eby:2015hyx} can be expressed in the form  
\begin{equation}
- \frac{dM}{dt} =  \frac{m_a \omega k}{192 \pi f_a^4}
\left| \int\!\! d^3 r \, e^{i \bm{k} \cdot \bm{r}} 
\Big[ \lambda_4 + \frac{\lambda_6 \psi^*\psi}{8 m_a f_a^2} + \ldots\Big] \psi^3 \right|^2,
\label{dN/dt-ESW}
\end{equation}
where $\bm{k}^2 = 9\omega^2-m_a^2$ and $\psi(r)$ is the common wavefunction of the condensed axions normalized so the number of axions is given in Eq.~\eqref{Naxion}. (An error by a factor of $4\pi^2$ was corrected in Ref.~\cite{Eby:2017azn}.) A result consistent with Eq.~\eqref{dN/dt-ESW} was also obtained  in Ref.~\cite{Mukaida:2016hwd}, where this loss mechanism was referred to as ``decay via spatial gradients''. Since $|\bm{k}| \approx \sqrt{8}\, m_a$, the loss comes from the small high-momentum tail of the wavefunction. In the case of the instanton potential, the expansion in powers of $\psi^*\psi$ in the integrand of the Fourier transform in Eq.~\eqref{dN/dt-ESW} can be summed to all orders in terms of a Bessel function \cite{Eby:2015hyx}. Eby et al.\ obtained a  result for the loss rate  in Eq.~\eqref{dN/dt-ESW} for the sine-Gordon model in the limit $\varepsilon_b \to 0$ \cite{Eby:2017azn}. Their exponential suppression factor is consistent with  Eq.~\eqref{dE/dt-epsilon}, with the argument differing by less than 2\%. Their result for the coefficient in the prefactor is $A = 2723$. It does not have the suppression factor of $\epsilon^2$ predicted in Ref.~\cite{Fodor:2009kf}.
 
\subsection{Collapse of Dilute Axion Star}
\label{sec:Collapse}

If a dilute axion star is embedded in a gas of unbound axions, thermalization can condense additional axions, increasing the mass of the axion star. If the mass $M$ of the axion star is near the critical value  $M_* \approx N_* m_a$, where $N_*$ is given by Eq.~\eqref{M-critical}, the condensation of additional axions can increase $M$ to above $M_*$. It will then be unstable to collapse. The fate of a collapsing dilute  axion star has not been definitively established. The possibilities for the remnant after the collapse include
\begin{itemize}
\item
a {\it black hole}, with Schwarzschild radius smaller than the critical radius $R_{99*}$ by about 15 orders of magnitude,
\item
a {\it dense axion star}, with radius smaller than $R_{99*}$ by about 7 orders of magnitude,
\item
a {\it dilute axion star}, with radius larger  than $R_{99*}$,
\item
{\it no remnant}, because of complete disappearance into scalar waves.
\end{itemize}
                 
Chavanis considered the possibility that the collapse of a dilute axion star produces a black hole  \cite{Chavanis:2016dab}. The axions were described by the GPP equations for $\psi$ and $\Phi$ given by Eqs.~\eqref{Newton-psiphi} with the truncated effective potential $V_\mathrm{eff}$ in Eq.~\eqref{Veff-approx}. The collapse was described by a Gaussian ansatz for the complex axion field $\psi(r,t)$ with a time-dependent radius $R(t)$. If the initial configuration is an unstable solution with mass $M$ near $M_*$, the time for collapse to $R=0$ scales as $(M-M_*)^{-1/4}$. Similar variational methods were used previously to study the collapse of gravitationally bound BEC's of bosons with a positive scattering length \cite{Harko:2014vya}.

Eby et al.\  followed Chavanis by describing the collapse of a dilute axion star using Eqs.~\eqref{Newton-psiphi} and a Gaussian ansatz for $\psi(r,t)$ with a time-dependent radius $R(t)$, but with $V_\mathrm{eff}$ given by the naive instanton effective potential in Eq.~\eqref{Veff-instanton}. The total energy for the Gaussian ansatz has a global minimum in $R$ that corresponds to a dense axion star. Eby et al.\  assumed that the collapse would somehow be stabilized at this smaller radius by inelastic reactions that produce relativistic axions, leaving a dense axion star as the remnant.

Helfer et al.\ studied the fate of spherically symmetric axion configurations by solving the full nonlinear classical field equations of general relativity for axions with the instanton potential \cite{Helfer:2016ljl}. Their initial condition was an oscillaton configuration for bosons with mass $m_a$ and no self-interactions and with a specified total mass $M$. By evolving the configurations in time, they found the possibilities for the remnant were a black hole, a dilute axion star, and no remnant. Their calculations were limited to the parameter region $4 \times 10^{-8} < G f_a^2 < 4 \times 10^{-2}$ and $0.03 < GM m_a <0.12$. The three possibilities for the remnant came  from separate regions of the plane of $G f_a^2$ versus $GM m_a$. Their results were consistent with the three regions meeting at a triple point given by $G f_a^2= 3.6 \times 10^{-3}$ and $GM m_a= 0.095$. The extrapolation of the results of Ref.~\cite{Helfer:2016ljl} to the tiny value of $G f_a^2$ for the QCD axion in Eq.~\eqref{Gf^2} implies that the only possibilities for the remnant are a black hole or  no remnant.

Levkov,  Panin, and Tkachev described the collapse of a dilute axion star above the critical mass by using the GPP equations \cite{Levkov:2016rkk}. The collapsing solutions approach a self-similar scaling limit with a singularity at a finite time $t_*$. To describe the behavior at later  times $t>t_*$, they used Eqs.~\eqref{Einstein-phiPhi} for  $\phi$ and $\Phi$ with the chiral potential $V(\phi)$, but with the  $\dot \Phi \dot \phi$ and $\Phi  \bm{\nabla}^2 \phi$ terms omitted. These equations predict multiple cycles of growth of the energy density near the center of the star followed by collapse. The collapse dramatically increases the energy density near the center, and it produces a burst of outgoing relativistic axion waves, which then depletes the energy density near the center. Levkov et al.\  found that after these multiple cycles, the remnant is gravitationally bound. They concluded that the remnant  must ultimately relax to a dilute axion star by gravitational cooling.

The collapse of the axion star followed by a burst of relativistic axions is an example of a {\it Bose nova}. A Bose nova was first produced by a group at JILA using a BEC of rubidium-85 atoms \cite{Donley:2001}. These atoms have a magnetic Feshbach resonance that can be used to control the scattering length $a$. The experiment began with a stable BEC with a positive scattering length in a trapping potential. The Feshbach resonance was used to suddenly reverse the sign of the scattering length, making the BEC unstable to collapse. The collapse of the BEC  produced a burst of high-energy atoms, analogous to the explosion produced by gravitational collapse in a supernova. The burst from the collapsing BEC was followed by its relaxation into a remnant BEC.

The collapse of a dilute axion star whose mass has increased to above the critical mass can be modeled more closely by experiments using a BEC of trapped atoms with a fixed negative scattering length $a$. The BEC in a trapping potential is stable unless the number $N$ of atoms exceeds a critical value. The first observations of the collapse of the BEC with fixed negative $a$ and well-controlled atom number were carried out by a group at Rice University using lithium-7 atoms \cite{Gerton:2014hsa}.

\subsection{Relativistic Corrections}
\label{sec:Relativistic}

The authors of Ref.~\cite{Visinelli:2017ooc} have argued that axions in dense axion stars may be too relativistic to use a nonrelativistic approach such as NREFT. Their evidence was not completely convincing except in the limit  $\omega \to 0$, in which the boson binding energy $\varepsilon_b$ approaches $m_a$. One way to test the accuracy of the nonrelativistic approximation is to compare the relativistic correction to the kinetic energy density, which is given by the second term in $\mathcal{T}_\mathrm{eff}$ in Eq.~\eqref{Teff-psi}, to the other terms in the energy density. For the critical dense axion star, the relativistic correction is small compared to the nonrelativistic kinetic energy density except near the center where the nonrelativistic kinetic energy density vanishes, as shown in Fig.~\ref{fig:Energy-r:axiton}. At the center, the relativistic correction  to the  kinetic energy density is smaller than the potential energy density by a factor of 0.36. This is large enough that one might question  the accuracy of the nonrelativistic approximation in the core of the critical dense axion star.

At the critical mass, the relativistic correction to the total kinetic energy is smaller than the total nonrelativistic kinetic energy by a factor of $-0.23$. As $M$ increases, the relativistic correction increases, but the nonrelativistic kinetic energy and the potential energy increase more rapidly, as shown in Fig.~\ref{fig:Energy-nhat:axiton}. Thus the accuracy of the nonrelativistic approximation may improve along the dense axion star branch as one moves away from the critical point.

\subsection{Higher Harmonics}
\label{sec:Harmonics}

The authors of Ref.~\cite{Visinelli:2017ooc} have pointed out that the harmonic approximation for $\phi(r,t)$ in Eq.~\eqref{phi-harmonic} may have large errors on the dense branch of axion stars. Accurate results may require taking into account the higher harmonics in the cosine series in Eq.~\eqref{phi-cosine}. The harmonic approximation for $\phi$ is equivalent to the harmonic approximation for $\psi$ in Eq.~\eqref{psi-periodic} along with the naive effective potential in Eq.~\eqref{V-naive}. There could therefore be significant  corrections from  higher harmonics to the properties of oscillons given in Eqs.~\eqref{criticalaxiton} and \eqref{axiton-scaling} and to the curves shown in Figs.~\ref{fig:RvsM-axiton}, \ref{fig:Energy-r:axiton}, and \ref{fig:Energy-nhat:axiton}.
 
One way to take the higher harmonics into account is to use a truncated odd cosine series for $\phi(r,t)$  as in Eq.~\eqref{phi-cosine} with a maximum frequency $n_\mathrm{max} \omega$, and to increase $n_\mathrm{max}$ until the results are stable. If gravity is important, one must also include even cosine series for the functions that determine the metric tensor. Such a procedure, with frequencies up to $12\, \omega$, was used  in Ref.~\cite{Helfer:2016ljl} to study the fate of axion stars.

In NREFT, the higher harmonics are integrated out in favor of local terms in the effective Hamiltonian. For example, the term $-(17/8) \lambda_4^2$ in the coefficient $v_3$ of the effective potential given in Eq.~\eqref{v3} takes into account the third harmonic of $\phi$, because the third diagram in Fig.~\ref{fig:3to3tree} has a virtual axion line whose invariant mass is approximately $3m_a$. Inside a dense axion star, the dimensionless number density $2\psi^*\psi/m_a f_a^2$ is large enough that the power series for $V_\mathrm{eff}(\psi^*\psi)$ cannot be truncated. In order to take into account higher harmonics, it is necessary to use approximations to the effective Hamiltonian that sum up additional terms with all powers of $\psi^*\psi$. An example is the sequence of improved effective potentials $V^{(k)}_\mathrm{eff}(\psi^*\psi)$ introduced in Ref.~\cite{Braaten:2018lmj}.

\subsection{Rotating Axion Stars}
\label{sec:Rotating}

Since an axion star is a superfluid, its angular momentum must be concentrated in vortices. Banik and Sikivie pointed out that the interactions between the vortices in an axion BEC are attractive, so the angular momentum in an axion star is concentrated in a single big vortex passing through the center \cite{Banik:2013rxa}. They also suggested  that instead of considering axion stars with definite values of the angular momentum, it might be more appropriate to consider definite values of the Legendre transform variable, which is an angular frequency.

Davidson and Schwetz have studied rotating dilute axion stars using the GPP equations given by Eqs.~\eqref{Newton-psiphi} with $V_\mathrm{eff}$ in Eq.~\eqref{Veff-approx} along with the simplifying assumption that the complex field $\psi(r, \theta,\phi)$ can be expressed as the product of a spherical harmonic $Y_{\ell m}( \theta,\phi)$ and a function of the radial coordinate $r$ \cite{Davidson:2016uok}. Similar approximations were used in Ref.~\cite{Hertzberg:2018lmt}. Sarkar, Vaz, and Wijewardhana studied rotating boson stars using the GPP equations with an additional term in Eq.~\eqref{Newton-psi} that takes into account frame dragging due to the rotation \cite{Sarkar:2017aje}.



\section{Acknowledgments}

We thank J.~Eby, L.~He, A.~Mohapatra, and S.~Raby  for useful discussions and comments. The research of E.B.\ was supported in part by the U.S.\ Department of Energy under grant DE-SC0011726 and by the U.S.\ National Science Foundation under grant PHY-1607190. The research of HZ was supported in part by the DFG Collaborative Research Center `Neutrinos and Dark Matter in Astro- and Particle Physics' (SFB 1258). EB  acknowledges the DFG cluster of excellence `Origin and Structure of the Universe' for support during a visit to Technische Universit\"at M\"unchen during which much of this review was written.

\bibliography{axionstars}

\end{document}